\documentclass[twocolumn]{aastex631}

\usepackage{graphicx}	
\usepackage{amsmath}	
\usepackage{subfigure}
\usepackage{natbib}
\usepackage{color}
\usepackage{tabularx}
\usepackage{longtable}
\usepackage{bm}
\usepackage{threeparttable}
\usepackage{enumerate}
\usepackage[figuresright]{rotating}
\usepackage[normalem]{ulem}
\usepackage{verbatim}

\newcommand{\code}[1]{{\texttt{#1}}}
\newcommand{\tess}{{\it TESS}}

\newcommand{\gaia}{{\it Gaia}}

\newcommand{\tar}{{TOI-5628}}

\begin{document}

\title{Stellar Multiplicity of M Dwarfs with Short-period Giant Planets, and the Characterization of TOI-5628Ab}

\correspondingauthor{Tianjun Gan}
\email{tianjungan@gmail.com}

\author[0000-0002-4503-9705]{Tianjun~Gan}
\affil{Department of Astronomy, Westlake University, Hangzhou 310030, Zhejiang Province, China}
\affil{Instituto de Astrof\'{i}sica de Canarias (IAC), E-38205 La Laguna, Tenerife, Spain}
\affil{Departamento de Astrof\'isica, Universidad de La Laguna (ULL), E-38206 La Laguna, Tenerife, Spain}

\author[0009-0005-6135-6769]{Alexandrine L'Heureux}
\affil{Institut Trottier de recherche sur les exoplan\`etes, D\'epartement de Physique, Universit\'e de Montr\'eal, Montr\'eal, QC H3C 3J7, Canada}

\author[0000-0001-9291-5555]{Charles Cadieux}
\affil{Observatoire de Gen\`eve, D\'epartement d’Astronomie, Universit\'e de Gen\`eve, Chemin Pegasi 51, 1290 Versoix, Switzerland}
\affil{Institut Trottier de recherche sur les exoplan\`etes, D\'epartement de Physique, Universit\'e de Montr\'eal, Montr\'eal, QC H3C 3J7, Canada}

\author[0000-0001-8317-2788]{Shude Mao} 
\affil{Department of Astronomy, Westlake University, Hangzhou 310030, Zhejiang Province, China}

\author[0000-0003-0987-1593]{Enric Pall\'{e}} 
\affil{Instituto de Astrof\'{i}sica de Canarias (IAC), E-38205 La Laguna, Tenerife, Spain}
\affil{Departamento de Astrof\'isica, Universidad de La Laguna (ULL), E-38206 La Laguna, Tenerife, Spain}

\author[0000-0002-6937-9034]{Sharon X. Wang} 
\affil{Department of Astronomy, Tsinghua University, Beijing 100084, People's Republic of China}

\author[0000-0002-3481-9052]{Keivan G. Stassun} 
\affil{Department of Physics and Astronomy, Vanderbilt University, 6301 Stevenson Center Ln., Nashville, TN 37235, USA}
\affil{Department of Physics, Fisk University, 1000 17th Avenue North, Nashville, TN 37208, USA}

\author[0000-0002-2532-2853]{Steve~B.~Howell} 
\affiliation{NASA Ames Research Center, Moffett Field, CA 94035, USA}

\author[0000-0002-3627-1676]{Benjamin V.\ Rackham}
\affil{Department of Earth, Atmospheric and Planetary Science, Massachusetts Institute of Technology, 77 Massachusetts Avenue, Cambridge, MA 02139, USA}
\affil{Kavli Institute for Astrophysics and Space Research, Massachusetts Institute of Technology, 77 Massachusetts Avenue, Cambridge, MA 02139, USA}

\author[0000-0002-0444-8502]{Steffani M. Grondin} 
\affil{Department of Astronomy \& Astrophysics, University of California San Diego, La Jolla, CA 92093, USA}

\author[0000-0003-1464-9276]{Khalid Barkaoui} 
\affil{Instituto de Astrof\'{i}sica de Canarias (IAC), E-38205 La Laguna, Tenerife, Spain}
\affil{Astrobiology Research Unit, Université de Liège, 19C Allée du 6 Août, 4000 Liège, Belgium}
\affil{Department of Earth, Atmospheric and Planetary Science, Massachusetts Institute of Technology, 77 Massachusetts Avenue, Cambridge, MA 02139, USA}


\author[0000-0002-0111-1234]{Luc Arnold}
\affiliation{Canada France Hawaii Telescope (CFHT) Corporation, UAR2208 CNRS-INSU, 65-1238 Mamalahoa Hwy, Kamuela 96743 HI, USA}

\author[0000-0003-3506-5667]{\'Etienne Artigau}
\affil{Institut Trottier de recherche sur les exoplan\`etes, D\'epartement de Physique, Universit\'e de Montr\'eal, Montr\'eal, QC H3C 3J7, Canada}
\affiliation{Observatoire du Mont-M\'egantic, Universit\'e de Montr\'eal, Montr\'eal, QC H3C 3J7, Canada}

\author[0000-0001-9892-2406]{Artem Burdanov} 
\affil{Department of Earth, Atmospheric and Planetary Science, Massachusetts Institute of Technology, 77 Massachusetts Avenue, Cambridge, MA 02139, USA}

\author[0000-0002-6523-9536]{Adam J.~Burgasser} 
\affil{Center for Astrophysics and Space Sciences, University of California, San Diego, 9500 Gilman Dr, La Jolla, CA 92093, USA}

\author[0000-0003-1963-9616]{Douglas A. Caldwell} 
\affil{NASA Ames Research Center, Moffett Field, CA 94035, USA}
\affil{SETI Institute, Mountain View, CA 94043, USA}

\author[0000-0002-5741-3047]{David R. Ciardi} 
\affiliation{NASA Exoplanet Science Institute, Caltech/IPAC, Mail Code 100-22, 1200 E. California Blvd., Pasadena, CA 91125, USA}

\author[0000-0001-6588-9574]{Karen A. Collins} 
\affil{Center for Astrophysics ${\rm \mid}$ Harvard {\rm \&} Smithsonian, 60 Garden Street, Cambridge, MA 02138, USA}

\author[0000-0003-4166-4121]{Neil J. Cook}
\affil{Institut Trottier de recherche sur les exoplan\`etes, D\'epartement de Physique, Universit\'e de Montr\'eal, Montr\'eal, QC H3C 3J7, Canada}

\author[0000-0001-5485-4675]{Ren\'e Doyon}
\affil{Institut Trottier de recherche sur les exoplan\`etes, D\'epartement de Physique, Universit\'e de Montr\'eal, Montr\'eal, QC H3C 3J7, Canada}
\affiliation{Observatoire du Mont-M\'egantic, Universit\'e de Montr\'eal, Montr\'eal, QC H3C 3J7, Canada}

\author[0000-0002-3937-630X]{Georgina Dransfield} 
\affil{Department of Astrophysics, University of Oxford, Denys Wilkinson Building, Keble Road, Oxford OX1 3RH, UK}
\affil{Magdalen College, University of Oxford, Oxford OX1 4AU, UK}

\author[0000-0002-4909-5763]{Akihiko Fukui} 
\affil{Komaba Institute for Science, The University of Tokyo, 3-8-1 Komaba, Meguro, Tokyo 153-8902, Japan}
\affil{Instituto de Astrof\'{i}sica de Canarias (IAC), E-38205 La Laguna, Tenerife, Spain}

\author[0000-0003-1462-7739]{Micha{\"e}l Gillon} 
\affil{Astrobiology Research Unit, Université de Liège, 19C Allée du 6 Août, 4000 Liège, Belgium}

\author[0000-0001-8923-488X]{Emmanuel Jehin} 
\affil{Space Sciences, Technologies and Astrophysics Research (STAR) Institute, Université de Liège, 19C Allée du 6 Août, 4000 Liège, Belgium}

\author[0000-0001-9087-1245]{Felipe Murgas} 
\affil{Instituto de Astrof\'{i}sica de Canarias (IAC), E-38205 La Laguna, Tenerife, Spain}
\affil{Departamento de Astrof\'isica, Universidad de La Laguna (ULL), E-38206 La Laguna, Tenerife, Spain}

\author[0000-0001-8511-2981]{Norio Narita} 
\affil{Komaba Institute for Science, The University of Tokyo, 3-8-1 Komaba, Meguro, Tokyo 153-8902, Japan}
\affil{Astrobiology Center, 2-21-1 Osawa, Mitaka, Tokyo 181-8588, Japan}
\affil{Instituto de Astrof\'{i}sica de Canarias (IAC), E-38205 La Laguna, Tenerife, Spain}

\author[0000-0003-4295-7313]{Federico R. Noguer} 
\affil{School of Earth and Space Exploration, Arizona State University, 781 E. Terrace Mall, Tempe, AZ 85287-6004, USA}

\author[0009-0009-5132-9520]{Howard M. Relles} 
\affil{Center for Astrophysics ${\rm \mid}$ Harvard {\rm \&} Smithsonian, 60 Garden Street, Cambridge, MA 02138, USA}

\author[0000-0002-1836-3120]{Avi Shporer} 
\affil{Department of Physics and Kavli Institute for Astrophysics and Space Research, Massachusetts Institute of Technology, Cambridge, MA 02139, USA}

\author[0000-0001-8536-0942]{Molly N. Simon} 
\affil{School of Earth and Space Exploration, Arizona State University, 781 E. Terrace Mall, Tempe, AZ 85287-6004, USA}

\author[0000-0002-0345-2147]{Abderahmane Soubkiou} 
\affil{Astrobiology Research Unit, Université de Liège, 19C Allée du 6 Août, 4000 Liège, Belgium}

\author[0000-0002-9807-5435]{Christopher A. Theissen} 
\affil{Center for Astrophysics and Space Sciences, University of California, San Diego, 9500 Gilman Dr, La Jolla, CA 92093, USA}

\author[0009-0008-2214-5039]{Mathilde Timmermans} 
\affil{School of Physics \& Astronomy, University of Birmingham, Edgbaston, Birmimgham B15 2TT, UK}
\affil{Astrobiology Research Unit, Université de Liège, 19C Allée du 6 Août, 4000 Liège, Belgium}

\author[0000-0002-5510-8751]{Amaury H. M. J. Triaud} 
\affil{School of Physics \& Astronomy, University of Birmingham, Edgbaston, Birmimgham B15 2TT, UK}

\author[0000-0001-7547-0398]{Robert T. Zellem} 
\affil{NASA Goddard Space Flight Center, 8800 Greenbelt Road, Greenbelt, MD 20771, USA}

\author[0000-0002-9350-830X]{S. Zúñiga-Fernández} 
\affil{Space Sciences, Technologies and Astrophysics Research (STAR) Institute, Université de Liège, 19C Allée du 6 Août, 4000 Liège, Belgium}



\begin{abstract}

Binary stars are ubiquitous, yet it remains unclear how wide-orbit stellar companions influence the formation of hot Jupiters, particularly around M dwarfs. Here, we first report the discovery of \tar Ab, a giant planet transiting a mid-type M dwarf ($M_\ast=0.36\pm0.02\ M_\odot$) every 4.34 days, accompanied by an associated white dwarf \tar B ($M_{\rm WD}=0.59\pm0.16\ M_\odot$) at a projected distance of about 2,500\,AU. Using TESS, ground-based photometry and SPIRou RVs, we constrain the planet radius to $0.74\pm0.04\ R_J$ and mass to $0.09\pm0.04\ M_J$, with a $3\sigma$ upper limit of $0.22\ M_J$. Building on this system, we further conduct a homogeneous systematic search for co-moving stellar companions with projected semi-major axis between 100 and 10,000\,AU around all M dwarfs with confirmed giant planets with periods smaller than 10 days and radii larger than 0.7 $R_J$, as well as a group of field M stars with stellar properties similar to the planet sample, based on the stellar kinematics from \textit{Gaia} DR3. We measure a stellar multiplicity of $34.2\pm9.5\%$ for M dwarfs hosting short-period giant planets, which is substantially higher than the fraction of $5.3\pm3.7\%$ for the field M stars by approximately a factor of 6. Our results suggest that wide-orbit stellar companions tend to promote the formation of short-period giant planets around M stars with masses $0.21 \leq M_\ast\leq 0.64\ M_\odot$, and high-eccentricity migration may play an important role in producing such systems.

\end{abstract}

\keywords{Stellar Multiplicity, M Dwarfs, Hot Jupiters, Stars: Individual (TIC 135100529)}


\section{Introduction} \label{sec:intro}

Stars are not alone: nearly half of local Sun-like stars were born in pairs or multiples \citep{Raghavan2010, Offner2023}, while low-mass M stars have a multiplicity, defined as the fraction of multiple-star systems in a sample, of about 20\% \citep{Ward-Duong2015,Winters2019, Offner2023, Clark2024}. Whether and how wide-orbit stellar companions influence the formation and evolution of planetary systems, particularly hot Jupiters with orbital period $P\leq 10$ days, remains an open question. 

High-eccentricity migration is one of the leading channels proposed to explain the existence of short-period hot Jupiter systems \citep{Dawson2018}. Under this framework, the orbit of a giant planet is excited through mechanisms like scattering \citep{Rasio1996,Ford2008,Chatterjee2008}, secular interactions \citep{Wu2011,Petrovich2015}, and Kozai-Lidov mechanism \citep{Kozai1962,Lidov1962,Naoz2016} driven by regular perturbation from an outer massive body such as a stellar companion. Subsequently, the orbit shrinks and circularizes through tidal dissipation \citep{hut81,Eggleton1998,Jackson2008}. 

Based on imaging surveys, \cite{Ngo2016} found that the multiplicity of Sun-like stars hosting hot Jupiters is $47\pm4\%$ at projected separations between 50 and 2000\,AU, approximately 2.9 times higher than the fraction of field stars within the same projected semi-major axis range \citep[see also][]{Ngo2015}. A similar finding was reported by \cite{Fontanive2019}, where the authors searched for wide binary companions among a mixed sample of giant planets and brown dwarfs, and determined a binary fraction of about 79\% between 20 and 10,000\,AU, roughly twice as high as that of field stars. In contrast, later work from \cite{Moe2021} proposed that wide-orbit stellar companions have either a negligible or negative effect on the formation of close-in giant planets orbiting G dwarfs by comparing the occurrence rates from transit and radial velocity (RV) surveys. 

However, equivalent statistical studies on hot Jupiters around M dwarfs, which provide sensitive tests of planet formation given their high mass ratios among planetary systems \citep{Laughlin2004,Ida2005,Kennedy2008,Liu2020,Burn2021}, are still lacking. This is primarily due to the low occurrence rate of hot Jupiters around M stars compared to their Sun-like counterparts \citep{Gan2023,Bryant2023,Glusman2026}. Nevertheless, the situation has now changed: the Transiting Exoplanet Survey Satellite \citep[TESS;][]{Ricker2015} has yielded around 40 detections of such systems over the last eight years \citep[e.g.,][]{Jordan2022,Gantoi530,Gan2023TOI4201,Kanodia2024,Hartman2024,Dransfield2026,Frensch2026}, making it possible to look into their demographics. 

Here, we first report the discovery and characterization of \tar Ab, a short-period gas giant around a mid-type (M3.0$\pm$1.0) M dwarf with a wide-orbit white dwarf (WD) companion. We next conduct a systematic search for co-moving stellar companions around all M dwarfs with confirmed short-period giant planets ($P\leq 10$~days, $R_p\geq 0.7\ R_J$), determine their multiplicity rate between 100 and 10,000 AU, and compare the result with that of field M stars. The paper is organized as follows: We describe the TESS and SPIRou RV observations in Section~\ref{sec:obs}, perform stellar characterization and a joint analysis in Section~\ref{sec:analysis}, carry out the population-level multiplicity study in Section~\ref{sec:multiplicity}, before we conclude in Section~\ref{sec:conclusion}.

\section{TESS Photometry and SPIRou RVs} \label{sec:obs}

\begin{figure*}
\centering
\includegraphics[width=0.9\textwidth]{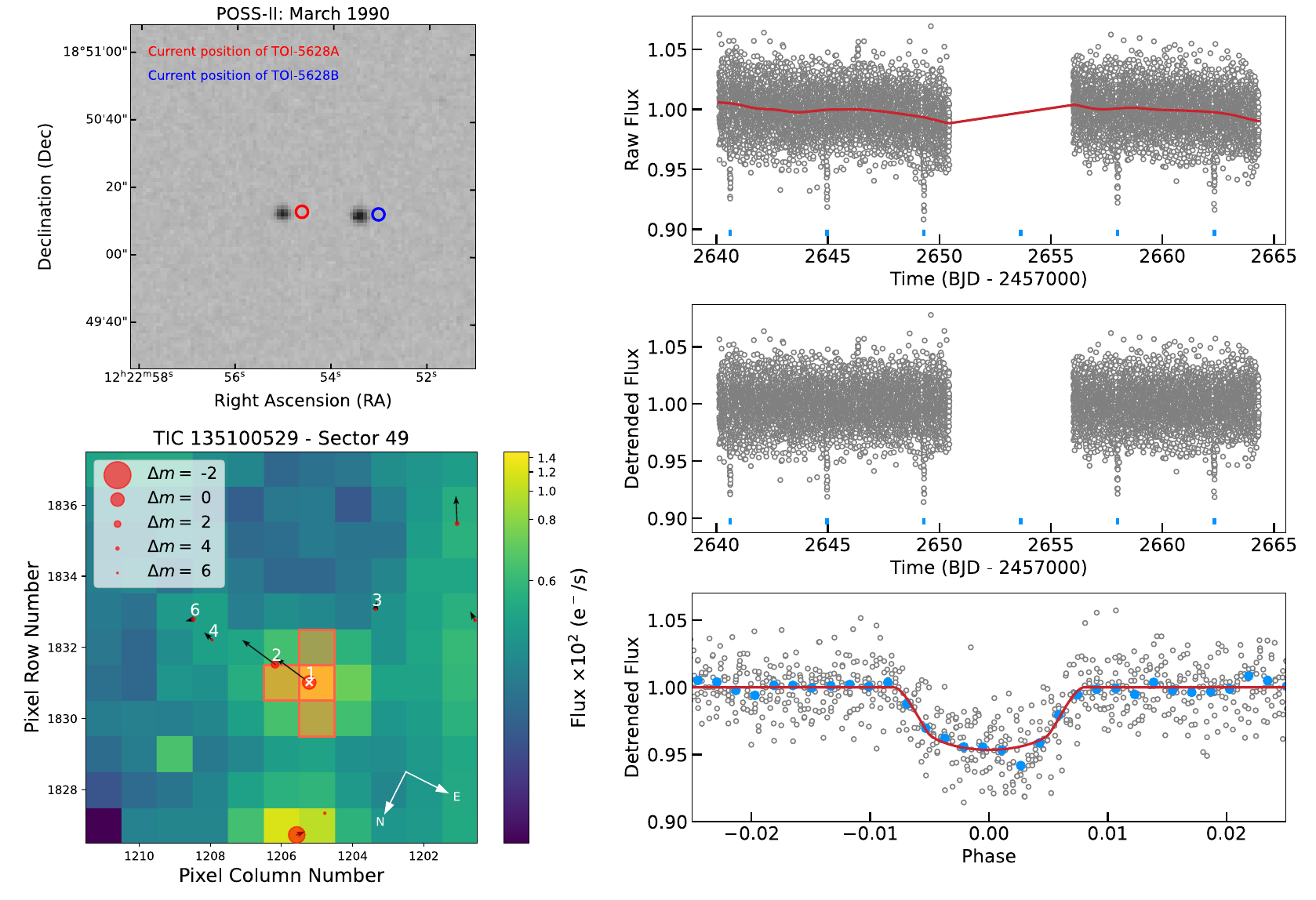}
\caption{Top left panel: The POSS-II image of \tar A and \tar B obtained in 1990. The red and blue circles are the current positions of two stars. Bottom left panel: The TPF plot of \tar A in Sector 49 with photometric aperture overlaid as the orange shaded region. \tar A is marked as a white cross. The black arrows represent the proper motions of every star. Right panel: The raw (top), detrended (middle), and phase-folded (bottom) TESS light curve. The transits are highlighted with blue ticks. The blue dots are the light curve binned to 10 minutes. The red curve is the best-fit model from the joint analysis. }
\label{FOV}
\end{figure*}

\tar A\ (LP 435-59) was monitored by TESS in Sector 49 during the Extended Mission under the two-minute cadence mode \citep{Ricker2015}. The observations spanned about 1 month, from UT 2022 March 1 to UT 2022 March 25, comprising 13404 exposures in total. The images were reduced by the Science Processing Operations Center \citep[SPOC;][]{Jenkins2016,Stumpe2012,Stumpe2014,Smith2012}. We made use of the simple aperture photometry (SAP) produced by SPOC without light dilution correction. After masking out the in-transit signals, we utilized the \code{celerite} algorithm \citep{Foreman2017} to fit a Gaussian Process (GP) model with a Mat\'{e}rn-3/2 kernel to the out-of-transit data to detrend the light curve. Figure~\ref{FOV} shows the POSS-II survey image of \tar\ taken in 1990 \citep{Lasker1996} alongside the TESS target pixel file (TPF) overlaid with the aperture used to extract the photometry, plotted using \code{tpfplotter} \citep{Aller2020}. The gravitationally-bounded WD \tar B (LP 435-58) located about 23 arcsec ($\sim 2500$~AU, $\Delta T=2.74$~mag) away from the target star appears in both images, which slightly contaminates the photometric aperture. The dilution effect is modeled and taken into account during the joint fit. The raw, detrended and phase-folded TESS light curves are shown in the right panel of Figure~\ref{FOV}. 

We collected 40 spectra\footnote{From programs led by T. Gan (24AS04, 24AD04 and 24AD09) and X. Hua (24BS05)} of \tar A between UT 2023 December 24 and UT 2024 December 21 using SPIRou \citep[SpectroPolarim\`etre InfraROUge;][]{Donati2020}, a near-infrared spectropolarimeter installed on the 3.6-m Canada-France-Hawaii Telescope (CFHT) with a spectral resolution of $R\approx 75,000$ and a wavelength range 0.95--2.35\,$\mu$m. Each spectrum was obtained from a 30-min exposure with \tar A in fibers A and B without using a simultaneous calibration lamp in fiber C to avoid contamination of the relatively faint science spectra by the bright Fabry-Pérot étalon. The observing set-up results in a median signal-to-noise ratio (SNR) of 23 at order 35 ($\sim1.75\,\mu$m) for each spectrum. We reduced the data with the standard reduction pipeline \code{APERO} v0.7.296 \citep{Cook2022}. To derive the RVs, we employed the line-by-line (LBL; v0.67.007) framework of \citet{Artigau2022}. Since \tar A is faint, we chose a high-SNR star of similar spectral type to build a robust template against which the per-line RV can be computed for \tar A. We find GJ~15B\footnote{Data publicly available through the Canadian Astronomical Data Centre (CADC): \url{https://www.cadc-ccda.hia-iha.nrc-cnrc.gc.ca}.} (M3.5V; \citealt{2013AJ....145..102L}) to provide the best match. The resulting RVs, with median precision of 17.6~m~s$^{-1}$, are shown in Figure~\ref{rv_plot}, and the data can be downloaded from ExoFOP\footnote{\url{https://exofop.ipac.caltech.edu/tess/target.php?id=135100529}}. Additional follow-ups, including ground-based photometric, spectroscopic, and imaging observations, are described in Appendix~\ref{ground_based_observations}. 

\begin{table}\scriptsize
    \centering
    \caption{Summary of stellar parameters for \tar  A}
    \begin{tabular}{lll}
        \hline\hline
        Parameter       &Value       &Ref. \\\hline
        \it{Main identifiers}                    \\
         TIC                     &$135100529$   &$\rm TIC\ V8^{[1]}$\\
         \gaia\ ID            &$3947322408155619456$ &\gaia\ DR3$^{[2]}$\\
         LP &435-59 &Luy63$^{[3]}$\\
         \it{Equatorial Coordinates} \\
         $\alpha_{\rm J2015.5}$    &12:22:54.71 &$\rm TIC\ V8$\\
         $\delta_{\rm J2015.5}$    &+18:50:12.14   &$\rm TIC\ V8$ \\
         \it{Photometric properties}\\
         $\tess$\ (mag)           &$14.473\pm0.006$   &$\rm TIC\ V8$  \\
         $B$ (mag) &$17.987\pm0.199$ &APASS$^{[4]}$\\
         $V$ (mag) &$16.902\pm0.076$ &APASS\\
         $G$ (mag)           &$15.571\pm0.003$   &\gaia\ DR3   \\
         $G_{\rm BP}$ (mag)           &$17.045\pm0.007$   &\gaia\ DR3   \\
         $G_{\rm RP}$ (mag)           &$14.386\pm0.004$   &\gaia\ DR3   \\
         $J$\ (mag)                    &$12.764\pm0.022$   &2MASS$^{[5]}$\\
         $H$\ (mag)                    &$12.165\pm0.024$   &2MASS \\
         $K$\ (mag)                    &$11.907\pm0.020$    &2MASS \\
         $W$1 (mag)                   &$11.770\pm0.022$   &WISE$^{[6]}$ \\
         $W$2 (mag)                   &$11.612\pm0.021$   &WISE \\
         $W$3 (mag)                   &$11.374\pm0.182$   &WISE \\
         \it{Astrometric properties}\\
         $\varpi$ (mas)              &$9.068\pm0.044$  &\gaia\ DR3  \\
         $\mu_{\alpha}\ ({\rm mas~yr^{-1}})$     &$-170.769\pm0.040$   &\gaia\ DR3   \\
         $\mu_{\delta}\ ({\rm mas~yr^{-1}})$     &$-28.445\pm0.042$   &\gaia\ DR3  \\
         \it{Stellar parameters} \\
         Distance (pc)                &$110.27\pm 0.54$  &\gaia\ DR3     \\
         Spectral Type &$\rm M3.0\pm 1.0$ & This work\\
         $M_{\ast}\ (M_{\odot})$ &$0.36\pm0.02$ &This work       \\
         $R_{\ast}\ (R_{\odot})$ &$0.35\pm0.02$ &This work       \\
         $T_{\rm eff}\ ({\rm K})$           &$3310\pm100$  &This work       \\
         $\rm [Fe/H]$  &$-0.10\pm 0.14$ &This work \\
         $P_{\rm rot}$\ (days) &$27.9\pm0.3$ &This work\\
         \hline
    \end{tabular}
    \begin{tablenotes}
    \item[1]  [1]\cite{Stassun2019tic}; [2]\cite{Gaia2023}; [3]\cite{Luyten1963}; [4]\cite{Henden2016}; [5]\cite{Cutri2003}; [6]\cite{wright2010}.
    \end{tablenotes}
    \label{starparam}
\end{table}

\section{Analysis}\label{sec:analysis}
\subsection{Stellar Characterization} \label{stellar_characterization}

\subsubsection{Properties of the Host Star \tar A}


We assigned a NIR spectral type using the SpeX Prism Library Analysis Toolkit \citep[SPLAT,][]{splat} by comparing the reduced spectrum of \tar A to single-star spectral standards in the IRTF Spectral Library \citep{Cushing2005, Rayner2009}.
The best-matching template is the M3V standard AD Leo, with the M3.5V standard GJ 273 providing only a marginally worse match.
We therefore adopt a NIR spectral type of M3.0$\pm$1.0 for TOI-5628A.
Applying the \citet{Mann2013} relation to this spectrum, we measured an iron abundance of $\mathrm{[Fe/H] = -0.10 \pm 0.14}$~dex.


According to the $K$-band apparent magnitude from 2MASS and stellar parallax from \gaia\ DR3, we compute the absolute magnitude $M_{K}$ of \tar A to be $6.695\pm0.023$. We thus obtain the stellar radius $R_\ast = 0.348\pm 0.020\ R_\odot$ through the $R_\ast$-$M_K$ empirical relation derived by \cite{Mann2015}, which is in agreement with the result $0.350\pm0.016\ R_\odot$ derived based on an independent relation between stellar angular diameter and color $V-K_s$ presented in \cite{Boyajian2014}. In addition, we also estimate the stellar mass from the empirical $M_K$ relations\footnote{\url{https://github.com/awmann/M_-M_K-}} of \cite{Mann2019}, leading to $M_\star = 0.356 \pm 0.018$~M$_\odot$. This is consistent with the value $0.363\pm0.015\ M_\odot$ we obtained using the polynomial relation reported by \cite{Benedict2016}. 

Finally, we estimate the stellar effective temperature $T_{\rm eff}$ via three ways. Firstly, we measure the bolometric magnitude of $M_{\rm bol}=9.424\pm 0.094$ mag, which corresponds to bolometric luminosity of $L_{\ast}=0.013\pm0.001\ L_{\odot}$, after accounting for the bolometric correction ${BC}_{K}=2.729\pm0.091$ calculated based on a relation with $V-J$ \citep{Mann2015}. We then obtain a $T_{\rm eff}=3324\pm53$~K through the Stefan-Boltzmann law. Similar results $3259\pm106$~K and $3320\pm83$~K are also derived when using the temperature-color relations reported in \cite{Pecaut2013} and \cite{Mann2015}, respectively.


As an independent determination of the basic stellar parameters, we perform an analysis of the broadband spectral energy distribution (SED) of the star together with the {\it Gaia\/} DR3 parallax \citep[with no systematic offset applied; see, e.g.,][]{StassunTorres:2021}, in order to determine an empirical measurement of the stellar radius, following the procedures described in \citet{Stassun:2016,Stassun:2017,Stassun:2018}. Where available, we pull the $JHK_S$ magnitudes from {\it 2MASS}, the W1--W3 magnitudes from {\it WISE}, the $G$, $G_{\rm BP}$, $G_{\rm RP}$ magnitudes from {\it Gaia}, the $ugrizy$ magnitudes from {\it Pan-STARRS}, and the FUV and NUV magnitudes from {\it GALEX}. Together, the available photometry spans the full stellar SED over the wavelength range of 0.4--10~$\mu$m (Figure~\ref{fig:sed}, top left).

For \tar A, we perform a fit using PHOENIX \citep{Husser:2013} stellar atmosphere models, with the free parameters being the effective temperature ($T_{\rm eff}$) and metallicity ([Fe/H]), as well as the extinction $A_V$, which we limit to maximum line-of-sight value from the Galactic dust maps of \citet{Schlegel:1998}. The resulting fit (Figure~\ref{fig:sed}) has a best-fit $A_V = 0.04 \pm 0.04$, $T_{\rm eff} = 3300 \pm 75$~K, [Fe/H] = $0.0 \pm 0.3$, with a reduced $\chi^2$ of 1.4. Integrating the (unreddened) model SED gives the bolometric flux at Earth, $F_{\rm bol} = 3.61 \pm 0.13 \times 10^{-11}$ erg~s$^{-1}$~cm$^{-2}$. Taking the $F_{\rm bol}$ and $T_{\rm eff}$ together with the {\it Gaia\/} parallax, gives the stellar radius, $R_\star = 0.359 \pm 0.016$~R$_\odot$, which agrees with the results from the empirical relation estimations within about $1\sigma$. We take the weighted values from different methods, and summarize the stellar parameters we adopted in Table~\ref{starparam}.



Moreover, we perform a frequency analysis on the long-term light curve from the Zwicky Transient Facility \citep[ZTF;][]{Bellm2019,Masci2019}. We apply generalized Lomb-Scargle periodogram \citep[GLS;][]{Zechmeister2009} to search for periodic modulations. Due to the faintness of the host star, we use the zr-band data with 597 measurements after excluding the bad points. Figure~\ref{ZTF} shows the periodogram, where a peak locates at $27.9\pm0.3$ days with false-alarm probability (FAP) below 0.1\%. A tentative baseline variation can also be seen in the TESS light curve (Figure~\ref{FOV}).

\subsubsection{Properties of the Wide-orbit WD \tar B}

We similarly fit the SED of the associated white dwarf companion \tar B, using the WD atmosphere models of \cite{Koester2010}, and enforcing the same $A_V$ as for \tar A (assuming it is at the same distance and line of sight). In that case, we obtain $T_{\rm eff} = 13600 \pm 500$~K, $\log g = 8.0 \pm 0.2$, and $R_{\rm WD} = 0.0117 \pm 0.0012$~R$_\odot$. The radius and $\log g$ together imply $M_{\rm WD} = 0.59 \pm 0.16$~M$_\odot$. The mass is consistent with the estimation $0.589\pm0.021\ M_\odot$ reported in \cite{Vincent2024}. According to \cite{Raddi2022}, the cooling age of \tar B is $0.529\pm0.011$~Gyr while the age of the progenitor is uncertain, ranging from 2.5 to 6.9 Gyr, based on the initial-to-final-mass relations from \cite{Catalan2008} and \cite{Cummings2018}.

While the broadband SED provides a good overall fit to a WD model (Figure \ref{fig:sed}, top right), spectroscopic confirmation is required to confirm the presence of a WD (and rule out degeneracies with other hot and compact sources, such as subdwarfs). In Section \ref{sec:shane}, we describe our observations and data reduction of \tar B using Shane/Kast, and present the resulting normalized spectrum in the bottom panel of Figure \ref{fig:sed}. In principle, a WD's $T_{\mathrm{eff}}$ and $\log g$ can be constrained directly through detailed fitting of Balmer line profiles using WD atmosphere models \citep[e.g.,][]{Koester2010, 2011ApJ...730..128T}, as these parameters strongly influence the depth and pressure-broadened shape of the lines. However, due to the low signal-to-noise ratio of the Shane/Kast spectrum, particularly in the blue region where the lower-order Balmer lines are located, as well as residual flux calibration uncertainties, a reliable quantitative atmospheric fit is not feasible (where the inferred uncertainties on $T_{\mathrm{eff}}$ and $\log g$ greatly exceed those obtained via the SED fit). Nevertheless, the spectrum clearly exhibits multiple broad Balmer absorption features at their expected wavelengths, providing robust confirmation of a DA WD. We therefore adopt the $T_{\mathrm{eff}}$ and $\log g$ values inferred from the SED fit above for the WD \tar B.

\subsection{Joint Analysis of Transits and RVs}\label{sec:joint-fit}

We utilize the \code{juliet} \citep{juliet} package to perform a joint analysis of all light curves together with the SPIRou RVs, which makes use of \code{batman} \citep{Kreidberg2015} to construct the light curve model as well as \code{radvel} \citep{Fulton2018} to fit the RVs. The posteriors are determined through dynamic nested sampling using \code{dynesty} \citep{Speagle2019}.

\begin{figure*}
\centering
\includegraphics[width=0.99\textwidth]{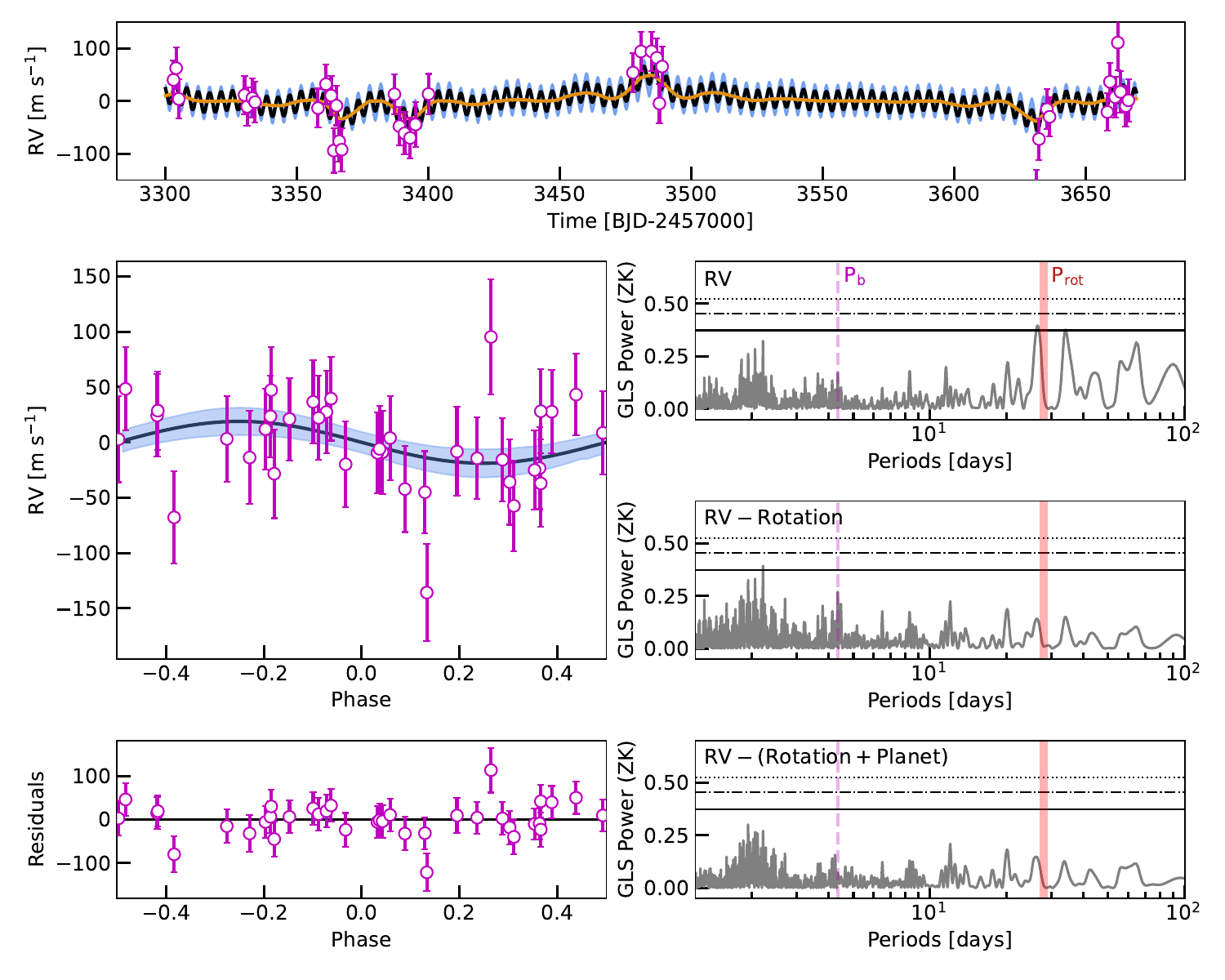}
\caption{Top panel: Time series of SPIRou RVs of \tar A along with the best-fit GP+Keplerian model. The orange curve represents the GP model. Middle left panel: The phase-folded RVs after subtracting the best-fit GP model, along with the Keplerian model shown as a black curve. In both panels, the blue shaded region marks the uncertainty at 68\% credibility interval. Bottom left panel: The residuals of fit, in $\rm m\ s^{-1}$. The presented RV error bars are the quadratic sum of the measurement uncertainties and the instrument jitter. Right panels: the generalized Lomb-Scargle periodograms of the RVs (top), RVs after excluding the GP model (middle) and the residuals after subtracting the GP and Keplerian models (bottom). The orbital period of the planet, and the stellar rotation period identified by the ZTF light curve are marked as vertical magenta and red lines. The false alarm probability levels of 10\%, 1\%, and 0.1\% are shown as horizontal solid, dot–dashed, and dotted lines.}
\label{rv_plot}
\end{figure*}

According to the frequency analysis of the long-term light curve from the Zwicky Transient Facility \citep[ZTF;][]{Bellm2019,Masci2019}, \tar A has a rotation period of $P_{\rm rot}=27.9\pm0.3$ days (see Section~\ref{stellar_characterization}). Such a periodic modulation also shows up in the generalized Lomb-Scargle periodogram \citep{Zechmeister2009} of SPIRou RVs (see Figure~\ref{rv_plot}) and tentatively presents in the TESS light curve. We therefore include a GP regression in the modeling with a rotation kernel formulated by \cite{Foreman2017}:
\begin{equation}
    k_{i,j}(\tau) = \frac{B}{2+C}e^{-\tau/L}\left[{\rm cos}\left(\frac{2\pi\tau}{P_{\rm rot}}\right) + (1+C) \right],
\end{equation}
where $B$ is the GP covariance amplitude, $C$ is a balance parameter between the periodic and the non-periodic terms, $\tau=|t_{i}-t_{j}|$ is the time-lag between two RV measurements $i$ and $j$, $L$ is the coherence timescale, and $P_{\rm rot}$ represents the stellar rotational period. 

We place wide non-informative priors on most parameters except for the TESS light dilution factor $D_{\rm TESS}$\footnote{The dilution factor is defined as $1/(1+A_{D})$, where $A_D$ is the flux ratio between all contaminating stars and the target \tar A.} and $P_{\rm rot}$ for which we adopt truncated-normal and normal priors (see Table~\ref{allposteriors}). For all ground-based photometry, we set the dilution factors to unity, as the target is well deblended. We adopt a quadratic limb-darkening law for the TESS photometry while a linear law for ground-based data for simplicity \citep{Kipping2013}. Given the short period of the planet that tidal force has circularized the orbit ($\tau_{\rm circ}\sim 0.41$~Gyr based on \citet{Jackson2008}), we opt to fix the orbital eccentricity to zero during the fit. For each instrument, we model an offset and a jitter to characterize additional white noise. We list the priors and the posteriors of key parameters in Table~\ref{allposteriors}. All photometric data along with the best-fit transit models are presented in Figures~\ref{FOV} and \ref{groundfit}. Figure~\ref{rv_plot} displays the RV time-series, the phase-folded RVs along with the GLS periodograms of the RVs, the RVs after subtracting the rotation signal, and the RVs after subtracting both the rotation and the planetary signal. The joint analysis reveals that the companion is a giant planet with a period of 4.34 days, a radius ratio of $0.2114\pm0.0008$ and a RV semi-amplitude of $\rm21.2\pm9.1\ m\ s^{-1}$, corresponding to a planet radius of $0.74\pm0.04\ R_J$ and a mass of $0.09\pm0.04\ M_J$ with a 3$\sigma$ upper limit of $0.22\ M_J$. We then rerun the fit but excluding the GP, and we obtain nearly the same RV semi-amplitude but a higher jitter. The GP+Keplerian model is slightly preferred over the Keplerian-only fit with a Bayesian evidence improvement of $\Delta\log Z=3$. However, we note that telluric correction is particularly challenging for faint targets when the Barycentric Earth Radial Velocity (BERV) approaches the stellar systemic velocity, causing stellar lines and telluric features from common species (e.g., H$_2$O and OH) to overlap (e.g., \citealt{Parc_2025, Osborn_2026, Srivastava_2026}). In our dataset, the target remains outside this ``BERV-crossing'' regime for most observations, with only six exposures in June 2024 (BJD$\sim2460485$) potentially affected. Consequently, we cannot rule out the possibility that part of the variability captured by the Gaussian Process component arises from residual telluric contamination.

In addition to the joint fit, we also carry out a transit timing variation (TTV) analysis to individual transits. We find no significant TTVs and all timings agree with the linear ephemeris within 5 mins.


\begin{table*}\scriptsize
    \centering
    {\renewcommand{\arraystretch}{1.05}
    \caption{Priors and the best-fit values of \tar Ab along with the 68\% credibility intervals in the joint fit. $\mathcal{N}$($\mu\ ,\ \sigma^{2}$) is a normal prior with mean $\mu$ and standard deviation $\sigma$, while $\mathcal{TN}$ ($\mu$, $\sigma^{2}$, $a$\ ,\ $b$) means a truncated normal prior ranging from $a$ to $b$. $\mathcal{U}$(a\ , \ b) represents a uniform prior between $a$ and $b$, while $\mathcal{LU}$(a\ , \ b) stands for a log-uniform prior between $a$ and $b$. }
    \begin{tabular}{lccr}
        \hline\hline
        Parameter       &Prior &Value    &Description\\\hline
        \it{Orbit parameters}\\
        $P$ (days)  &$\mathcal{U}$ ($1$\ ,\ $10$)  &$4.3416612\pm0.0000015$
        &Orbital period.\\
        $T_{c}$ (BJD-2457000) &$\mathcal{U}$ ($2637$\ ,\ $2643$) &$2640.6193\pm0.0002$ &Mid-transit time.\\
        $r_1$ &$\mathcal{U}$ ($0$\ ,\ $1$) &$0.6012\pm0.0131$ &Parametrization for p and b$^{[1]}$.\\
        $r_2$ &$\mathcal{U}$ ($0$\ ,\ $1$) &$0.2114\pm0.0008$ &Parametrization for p and b.\\
        $e$ &Fixed &0 &Orbital eccentricity.\\
        $\omega$ (deg) &Fixed &90 &Argument of periapsis.\\
        $K$ (m~s$^{-1}$) &$\mathcal{U}$ ($0$\ ,\ $1000$) &$21.2\pm9.1$ &RV semi-amplitude.\\
        $\mu_{\rm SPIRou}$ (m~s$^{-1}$) &$\mathcal{U}$ ($-30000$\ ,\ $-10000$) &$-20538.5\pm 9.7$ &RV offset.\\
        $\sigma_{\rm SPIRou}$ (m~s$^{-1}$) &$\mathcal{U}$ ($0$\ ,\ $100$) &$40.5\pm 6.0$ &RV jitter.\\\hline
        \it{Stellar parameter}\\
        $\rho_{\ast}$ (kg~m$^{-3}$) &$\mathcal{LU}$ ($10^{3}$\ ,\ $10^{6}$) &$11507\pm301$ &Stellar density of the host star.\\\hline
        \it{Dilution factors}\\
        $D_{\rm TESS}$ &$\mathcal{TN}$ ($0.9$, $0.1^{2}$, $0$\ ,\ $1$) &$ 0.879\pm0.027$ & \\
        $D_{\rm Ground}$ &Fixed  &1  \\\hline
        \it{GP hyperparameters}\\
        $B$ &$\mathcal{LU}$ ($10^{-6}$\ ,\ $10^{6}$) &$521.22^{+325.76}_{-516.65}$ & \\
        $C$ &$\mathcal{LU}$ ($10^{-6}$\ ,\ $10^{6}$) &$1.32^{+61.86}_{-1.30}$ & \\
        $L$ &$\mathcal{LU}$ ($10^{-2}$\ ,\ $10^{2}$) &$11.91^{+24.81}_{-11.43}$ & \\
        $P_{\rm rot}$ &$\mathcal{N}$ ($27.9$\ ,\ $2.0^{2}$) &$27.36\pm1.56$ & \\\hline
        \it{Derived parameters}\\
        $R_{p}/R_{\ast}$ &$\cdots$ &$0.2114\pm0.0008$ &Scaled planet radius.\\
        $a/R_{\ast}$ &$\cdots$ &$22.5498\pm0.1949$ &Scaled semi-major axis.\\
        $b$ &$\cdots$ &$0.4018\pm0.0196$ &Impact parameter.\\
        $i$ (deg) &$\cdots$ &$88.98\pm0.06$ &Orbital inclination.\\\hline
        \it{Physical parameters}\\
        $R_{p}$ ($R_{J}$) &$\cdots$ &$0.74\pm 0.04$ &Planet radius.\\
        $M_{p}$ ($M_{J}$) &$\cdots$ &$0.09\pm 0.04$ &Planet mass.\\
        $\log g_{\rm p}$ (cgs) &$\cdots$ &$2.61\pm0.27$ &Planet surface gravity.\\
        $a$ (AU) &$\cdots$ &$0.0368\pm 0.021$ &Semi-major axis.\\
        $\rho_{p}$ (g~cm$^{-3}$) &$\cdots$ &$0.29\pm 0.14$ &Planet density.\\
        $T_{\rm eq}^{[2]}$ (K) &$\cdots$ &$492\pm 15$ &Equilibrium temperature.\\
        \hline
    \label{allposteriors}    
    \end{tabular}}
    \begin{tablenotes} 
       \item[1]  [1] The parameters p and b are the companion-to-star radius ratio and the impact parameter \citep{Espinoza2018}. [2] We assume albedo $A_{B}=0$.
    \end{tablenotes}
\end{table*}

\section{Multiplicity of M Dwarfs with Short-period Giant Planets between 100 and 10,000 AU}\label{sec:multiplicity}

Putting the \tar Ab system into a broader context, it turns out that wide-orbit stellar companions around M dwarfs harboring short-period giant planets are not rare, and several cases have been reported \citep{Jordan2022,Canas2023,Bernabo2024,Bryant2024,Hartman2024,Kanodia2024,Reji2025,Frensch2026}. However, the detections of these stellar companions relied on different methods such as high-angular-resolution speckle imaging, adaptive optics observations, and Gaia DR2 or DR3 kinematics \citep{Gaia2018,Gaia2023}. Hence, the sample is heterogeneous and not appropriate for demographics. More importantly, it is yet unclear if wide-orbit stellar companions promote or impede the formation of hot Jupiters around low-mass stars. To address this question, we carry out a homogeneous systematic search for co-moving stellar companions with projected distance between 100 and 10,000\,AU around M dwarfs hosting short-period giant planets based on the stellar kinematical information provided in Gaia DR3. We measure the multiplicity of M dwarfs hosting giant planets with orbital periods below 10 days and radius above $0.7\ R_J$, and then compare our finding with that of the field M dwarfs. We outline the construction of the planet sample and our search for wide-orbit companions in Section~\ref{sample_construction}, and we present the multiplicity result of field M dwarfs within the same projected semi-major range in Section~\ref{comparison_to_field} before summarizing in Section~\ref{comparison_two_samples}.

\subsection{Planet Sample Construction and Co-moving Companion Search}\label{sample_construction}

We retrieve the list of hot Jupiters around M dwarfs from the NASA Exoplanet Archive \citep{Akeson2013,Christiansen2025} on UT 2026 May 1 using the following selection criteria:
\begin{enumerate}[(1)]
    \item Stellar effective temperature $T_{\rm eff}\leq 4,000$~K;
    \item Stellar mass $M_{\ast}\leq 0.65\ M_\odot$;
    \item Orbit period $P\leq 10$~days;
    \item Planets with radius $R_{p}\geq 0.7\ R_J$;
    \item Planets with mass determination or upper limits from RV observations.
\end{enumerate}

We emphasize that we define hot Jupiters as planets with orbital periods $P\leq 10$~days and radii $R_p\geq0.7\ R_J$ throughout this work. After including \tar Ab reported in this work, our planet sample contains a total of 38 systems. The host stars in the resulting sample have effective temperatures $3000\lesssim T_{\rm eff}\lesssim 4000$~K, stellar masses $0.21\lesssim M_\ast \lesssim 0.64\ M_\odot$, stellar distances $70\lesssim d\lesssim 400$ pc and TESS-band magnitude $12.5\lesssim T{\rm mag}\lesssim 15.0$. We record their Gaia DR3 ID, Right Ascension ($\alpha$), Declination ($\delta$) and distance. 

To evaluate the multiplicity, we follow the methodology proposed in \cite{ElBadry2021} to identify co-moving companions with analogous proper motions and parallaxes. Specifically, we perform a cone search centered at the ($\alpha$, $\delta$) coordinates of each system with a radius boundary extending to 15,000 AU. We conservatively expand the projected radius limit here to account for the uncertainties on stellar position and distance reported in the literature. For every neighboring source, we examine their kinematical properties and we regard a companion as a co-moving star if it satisfies the conditions below:

\begin{enumerate}[(1)]
\item The projected distance between the host and the companion is between 100 and 10,000 AU:
\begin{equation}
    100\ {\rm AU} \leq \frac{1000}{\varpi_1\ {\rm mas^{-1}}}\times \rho_{\rm dist}\leq 10,000\ {\rm AU},
\end{equation}
where $\varpi_1$ is the parallax of the target star, and $\rho_{\rm dist}$ is the projected separation distance between the target and the companion in units of arcsec, both taken from Gaia DR3;
\item The proper motion difference between the target and the companion is within 5 $\rm km\ s^{-1}$ with a $2\sigma$ tolerance, or equivalently:
\begin{equation}
    |\Delta \mu| \leq 1.05\ {\rm mas\ yr^{-1}} \times (\varpi_{1}+2\sigma_{\varpi_1})\ {\rm mas^{-1}},
\end{equation}
where $|\Delta \mu|=\sqrt{(\mu_{\alpha_1}-\mu_{\alpha_2})^{2}+(\mu_{\delta_1}-\mu_{\delta_2})^{2}}$, $\mu_{\alpha_1}$, $\mu_{\delta_1}$, $\mu_{\alpha_2}$, $\mu_{\delta_2}$ are the proper motions of the target and companion along the direction of Right Ascension and Declination, $\varpi_1$ and $\sigma_{\varpi_1}$ are the parallax of the target star and its uncertainty;

\item The parallaxes of the target and the companion are consistent within $5\sigma$:
\begin{equation}
    |\Delta \varpi| \leq 5 \sqrt{\sigma^{2}_{\varpi_1}+\sigma^{2}_{\varpi_2}},
\end{equation}
where $|\Delta \varpi|=|\varpi_{1}-\varpi_{2}|$, $\sigma_{\varpi_1}$ and $\sigma_{\varpi_2}$ are the parallax uncertainties of the target and the companion. While \cite{ElBadry2021} proposed a $2\sigma$ level parallax consistency, we loosen the constraint to $5\sigma$ following \cite{Rice2022warmjupiter}, but we emphasize that the choice here will not change the conclusion.
\end{enumerate}

Among the total sample of 38 systems, our systematic search leads to the detections of 14 co-moving stellar companions in 13 systems, including the identification of a stellar companion TIC 46432937B that was not reported in the previous work \citep{Hartman2024}. Although two co-moving companions of TOI-6034 were alerted by the search pipeline, the second object (Gaia DR3 2270399053598645120 at 8661 AU) has large uncertainties on its kinematics, and it is likely a background star according to its parallax. However, this ambiguous companion will not affect our statistics as we are investigating the multiplicity. We summarize the target and companion information in Table~\ref{search_results}. 

Our searching result can be simply converted to a multiplicity of ${P({\rm Comp|HJ})}=34.2\pm9.5\%$ for M dwarfs harboring hot Jupiters, where the uncertainty comes from Poisson noise. We note that the multiplicity derived here should be treated as a lower limit as we do not consider the detection sensitivity of Gaia. For example, the planetary system HATS-74Ab has a bound stellar neighbor denoted as HATS-74B at $238.4\pm3.9$~AU found by Keck/NIRC2 AO observations \citep{Jordan2022}, but its kinematical information is not recorded in Gaia DR3 due to its faintness and small projected separation distance. 

In addition to the radius-based hot Jupiter sample ($R_p\geq 0.7\ R_J$), we repeat our analysis using a mass-based planet sample. We filter out hot Jupiters with mass above $0.3\ M_J$, leading to a total of 31 systems except for TOI-762, TOI-3629, TOI-3757, TOI-3984, TOI-5628, TOI-5634 and TOI-6894 listed in Table~\ref{search_results}. Among them, a total of 9 systems have co-moving stellar companions, yielding a multiplicity of $P({\rm Comp|HJ})=29\pm10\%$. This is consistent with our previous estimation within $1\sigma$. Therefore, our result is not sensitive to the definition of hot Jupiters.

\begin{sidewaystable*}
    \vspace{30em}
    \centering
    \fontsize{6pt}{8pt}\selectfont
    {\renewcommand{\arraystretch}{1.1}
    \caption{Kinematical Properties of M dwarfs Hosting Hot Jupiters and their Co-moving Companions from the Systematic Search}
    \begin{tabular}{llccclccccc}
        \hline\hline
        Target &Target ID$^{[1]}$ &$\mu_{\alpha_1}$  &$\mu_{\delta_1}$   &$\varpi_1$ &Companion ID &$\mu_{\alpha_2}$  &$\mu_{\delta_2}$   &$\varpi_2$ &$\rho_{\rm dist}$ &Projected Distance\\
        & &$[{\rm mas\ yr^{-1}}]$ &$[{\rm mas\ yr^{-1}}]$ &$[{\rm mas}]$ & &$[{\rm mas\ yr^{-1}}]$ &$[{\rm mas\ yr^{-1}}]$ &$[{\rm mas}]$ &[arcsec] &[AU] \\\hline
        HATS-6 &2966680597368750720 &$-2.568\pm0.015$ &$7.371\pm0.018$ &$5.892\pm0.019$\\
        HATS-71 &4710594412266148352 &$79.128\pm0.029$ &$-26.985\pm0.027$ &$7.112\pm0.024$ \\
        HATS-74 &3545653561942122368 &$-38.992\pm0.043$ &$39.578\pm0.034$ &$3.425\pm0.042$ \\
        HATS-75 &5082914338199586560 &$12.716\pm0.013$ &$-1.687\pm0.021$ &$5.061\pm0.021$ \\
        Kepler-45 &2053562475706063744 &$4.646\pm0.036$ &$-21.589\pm0.036$ &$2.598\pm0.033$\\
        NGTS-1 &4821739369794767744 &$-31.887\pm0.017$ &$-41.077\pm0.021$ &$4.594\pm0.017$ \\
        TIC 46432937 &2984391358868786816 &$-13.365\pm0.013$ &$36.962\pm0.012$ &$11.031\pm0.013$ &2984391358868786560 &$-14.133\pm0.020$ &$37.209\pm0.018$ &$11.015\pm0.020$ &39.709 &3599.865 \\
        TOI-519 &5707485527450614656 &$-41.959\pm0.029$ &$29.074\pm0.027$ &$8.681\pm0.037$ \\
        TOI-530 &3353218995355814656 &$13.622\pm0.027$ &$-62.524\pm0.023$ &$6.770\pm0.028$ \\
        TOI-762 &5362352744504000256 &$-159.174\pm0.020$ &$-24.780\pm0.020$ &$10.118\pm0.023$ &5362352744496315264 &$-157.370\pm0.126$ &$-24.379\pm0.123$ &$9.792\pm0.144$ &3.290 &325.151\\
        TOI-2379 &6521531466699512064 &$16.575\pm0.014$ &$1.750\pm0.018$ &$4.755\pm0.018$ \\
        TOI-2384 &4699702272124241152 &$9.257\pm0.061$ &$-52.662\pm0.053$ &$5.322\pm0.044$ &4699702272123475328 &$8.013\pm0.623$ &$-51.041\pm0.255$ &$5.052\pm0.167$ &0.815 &153.055\\
        TOI-3235 &6107144260251920000 &$-170.503\pm0.028$ &$-64.264\pm0.023$ &$13.781\pm0.027$ \\
        TOI-3288 &6685431748042347776 &$-6.566\pm0.022$ &$-20.080\pm0.014$ &$4.969\pm0.025$ &6685431748040148992 &$-6.719\pm0.114$ &$-20.952\pm0.073$ &$5.384\pm0.120$ &2.131 &428.902\\
        TOI-3629 &2881820324294985856 &$185.707\pm0.012$ &$1.010\pm0.012$ &$7.667\pm0.017$ \\
        TOI-3714 &178924390478792320 &$19.826\pm0.025$ &$-70.762\pm0.020$ &$8.839\pm0.022$ &178924390476838784 &$18.805\pm0.252$ &$-70.672\pm0.205$ &$8.847\pm0.227$ &2.659 &300.800\\
        TOI-3757 &996878131494639488 &$-9.032\pm0.021$ &$-43.128\pm0.018$ &$5.598\pm0.021$ \\
        TOI-3984 &1291955578869575552 &$-48.951\pm0.016$ &$42.647\pm0.020$ &$9.184\pm0.019$ &1291955574574621056 &$-50.379\pm0.163$ &$40.950\pm0.212$ &$9.057\pm0.202$ &3.292 &358.448\\
        TOI-4201 &2997312063605005056 &$11.731\pm0.017$ &$6.053\pm0.018$ &$5.291\pm0.019$ \\
        TOI-4666 &4855422771071903232 &$4.122\pm0.014$ &$-22.135\pm0.019$ &$6.488\pm0.016$\\
        TOI-4860 &3571038605366263424 &$-177.238\pm0.039$ &$-5.430\pm0.033$ &$12.377\pm0.034$ \\
        TOI-5007 &5833386583912265728 &$-13.515\pm0.026$ &$-29.914\pm0.025$ &$4.909\pm0.022$ \\
        TOI-5205 &1842656663520849024 &$41.678\pm0.022$ &$52.074\pm0.024$ &$11.464\pm0.026$ \\
        TOI-5292 &2610422023759160576 &$-8.733\pm0.041$ &$-6.862\pm0.039$ &$2.790\pm0.042$ &2610422028054589824 &$-9.300\pm0.171$ &$-7.163\pm0.169$ &$2.766\pm0.178$ &9.400 &3369.006\\
        TOI-5293 &2640121486388076032 &$-17.131\pm0.031$ &$0.487\pm0.024$ &$6.165\pm0.028$ &2640121482094497024 &$-16.497\pm0.152$ &$-0.079\pm0.119$ &$6.274\pm0.133$ &3.538 &573.956 \\
        TOI-5344 &52359538285081728 &$40.325\pm0.028$ &$-22.194\pm0.020$ &$7.305\pm0.023$ \\
        TOI-5349 &58372904816938240 &$37.203\pm0.029$ &$-19.559\pm0.024$ &$5.262\pm0.026$ \\
        TOI-5573 &1023967108706495360 &$-3.586\pm0.034$ &$-56.841\pm0.034$ &$5.335\pm0.038$ \\
        \textbf{TOI-5628} &3947322408155619456 &$-170.769\pm0.040$ &$-28.445\pm0.042$ &$9.068\pm0.044$ &3947322403860036352 &$-170.202\pm0.094$ &$-28.622\pm0.088$ &$8.972\pm0.097$ &22.674 &2500.436\\
        TOI-5634 &3979511431397114752 &$-60.799\pm0.056$ &$-25.488\pm0.052$ &$3.090\pm0.059$ &3979511431397963008 &$-60.802\pm0.213$ &$-25.625\pm0.207$ &$2.899\pm0.234$ &3.815 &1234.784 \\
        TOI-5688 &1363205856494897024 &$-0.910\pm0.034$ &$64.930\pm0.036$ &$4.400\pm0.029$ &1363205856494896896 &$-0.693\pm0.087$ &$64.965\pm0.089$ &$4.282\pm0.073$ &4.885 &1110.252\\
        TOI-5916 &1741429438313012608 &$-8.816\pm0.039$ &$-25.598\pm0.040$ &$5.097\pm0.042$ \\
        TOI-6034 &2270404997834401664 &$-2.430\pm0.020$ &$-5.109\pm0.021$ &$8.512\pm0.016$ &2270407952771897472 &$-2.238\pm0.016$ &$-5.556\pm0.017$ &$8.528\pm0.014$ &40.068 &4707.269\\
        & & & & &2270399053598645120 &$1.470\pm1.820$ &$2.964\pm1.796$ &$4.637\pm1.346$ &73.725 &8661.279\\
        TOI-6303 &239050153051494272 &$33.035\pm0.021$ &$-35.352\pm0.018$ &$6.587\pm0.019$ \\
        TOI-6330 &407530931116600320 &$-36.516\pm0.032$ &$-45.080\pm0.025$ &$6.967\pm0.027$ \\
        TOI-6383 &473934011733049856 &$41.023\pm0.035$ &$-51.299\pm0.032$ &$5.847\pm0.031$ &473934007432818688 &$41.016\pm0.167$ &$-51.283\pm0.155$ &$6.099\pm0.158$ &18.151 &3104.601\\
        TOI-6894 &3917278287286247808 &$-146.897\pm0.056$ &$22.227\pm0.053$ &$13.684\pm0.053$ \\
        TOI-7149 &1200751810900182400 &$-7.698\pm0.039$ &$12.882\pm0.041$ &$7.860\pm0.046$ \\
        \hline
    \label{search_results}    
    \end{tabular}}
    \begin{tablenotes}
       \item[1]  [1]\ Gaia DR3 ID \citep{Gaia2023}.
    \end{tablenotes}
\end{sidewaystable*}

\subsection{Multiplicity of Field M Dwarfs}\label{comparison_to_field}

We adopt two approaches to measure the multiplicity of field M dwarfs within the same projected distance range (100-10,000 AU), which deliver consistent results. 

\begin{figure}
\centering
\includegraphics[width=0.49\textwidth]{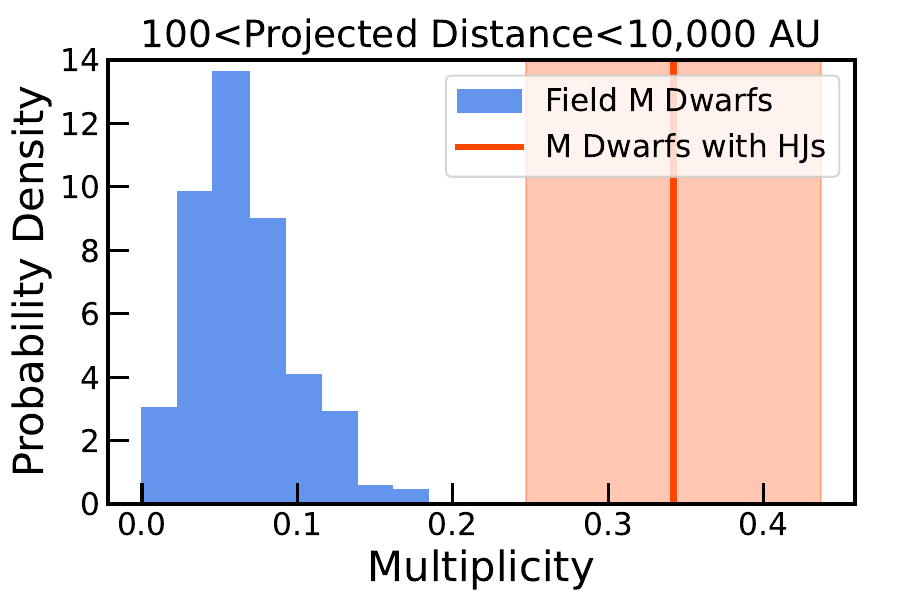}
\caption{Comparison of the stellar multiplicity of M dwarfs with hot Jupiters and field M stars at projected separations between 100 and 10,000 AU. The multiplicity of the planet sample, $P({\rm Comp|HJ})=34.2\pm9.5\%$, is indicated as a vertical red line with the shaded region marking the 1$\sigma$ uncertainty. The multiplicity of field M dwarfs, $P({\rm Comp})=5.3\pm3.7\%$, is shown as a histogram based on the search for 1000 sets of randomly selected field M stars with properties matched to those of the planet sample.}
\label{field_multiplicity_MC}
\end{figure}

Based on the cool dwarf list \citep[cooldwarfs\_v8;][]{Muirhead2018} embedded in the TESS input Catalog v8.2 \citep[TIC;][]{Stassun2019tic}, we filter out field M dwarfs that have similar brightness ($T{\rm mag}$), distance, effective temperature and mass as our planet sample. We require the field M dwarfs in our sample have flux contamination below 0.1, ensuring relatively accurate stellar properties. A total of 546,681 field M dwarfs are left. We employ a Monte Carlo method by randomly choosing 38 field stars, the same size of our planet sample, rerunning the companion search pipeline and computing the multiplicity $P({\rm Comp})$ in the same way introduced in Section~\ref{sample_construction}. We repeat the same process for 1000 times and record the multiplicity value measured in each trial. Figure~\ref{field_multiplicity_MC} shows the multiplicity distribution of field M dwarfs. We take the median and standard deviation of the distribution as the final multiplicity and its corresponding uncertainty, leading to $P({\rm Comp})=5.3\pm3.7\%$. Without adding the aforementioned flux contamination criterion that potentially excludes high-mass bright co-moving companions, we have 824,253 field M dwarfs left. We repeat the same analysis, and we derive a slightly higher multiplicity of $P({\rm Comp})=8.6\pm 4.5\%$. A similar result here is expected as the majority of wide-orbit co-moving companions around M dwarfs are skewed toward faint late-type M stars with smaller masses, hence having low light contamination ratios \citep{Ward-Duong2015,El-Badry2019,Offner2023}. 

In addition, we also explore the multiplicity of field M dwarfs with the help of results from an infrared adaptive optics survey on local M dwarfs within 15 pc \citep{Ward-Duong2015}, which derived the companion frequency in six different projected separation bins between 3 and 10,000 AU. We retrieve the observational results in Figure 16 of \cite{Ward-Duong2015}, and fit a standard Gaussian function to approximate the data. We obtain

\begin{equation}\label{eq:field_multiplicity}
    \frac{dP}{d\log a}\Delta\log a=0.09\times e^{\frac{(\log a+0.395)^{2}}{2\times2.13^{2}}},
\end{equation}
where $\Delta\log a= (4-\log 3)/6=0.587$ is the binning size used in \cite{Ward-Duong2015}. To constrain the upper limit on the multiplicity of field M dwarfs between 100 and 10,000 AU, we integrate Equation~\ref{eq:field_multiplicity} between $\log a_{\rm min}=2$ and $\log a_{\rm max}=4$:

\begin{equation}
    P({\rm Comp}) \leq  \int_{\log a_{\rm min}}^{\log a_{\rm max}} \frac{dP}{d\log a} \,d\log a \approx9.0\%,
\end{equation}
which agrees with the first-method measurement $P({\rm Comp})=5.3\pm3.7\%$ within $1\sigma$. The result 9.0\% from the second approach is, in fact, the companion frequency, quantifying the total number of companions within a sample, which is intrinsically equal to or higher than the multiplicity that quantifies the number of multiple-star systems within a sample. Moreover, it has been corrected for incompleteness (see Section 5.6 in \citealt{Ward-Duong2015}). Both reasons explain why it is slightly higher than the first estimate. Consequently, we adopt the first multiplicity value in the following analysis for consistency. 

\subsection{Multiplicity Comparison between the Planet and Field M Dwarf Sample}\label{comparison_two_samples}

Based on $P(\rm Comp|HJ)$ and $P({\rm Comp})$, one can compute the probability ratio $P({\rm HJ|Comp})/P(\rm HJ)$ through the Bayes' theorem:

\begin{equation}
    \frac{P({\rm HJ|Comp})}{P({\rm HJ})}=\frac{P(\rm Comp|HJ)}{P({\rm Comp})}\approx6.5
\end{equation}
Our result indicates that a wide-orbit stellar companion between 100 and 10,000\,AU enhances the formation efficiency of hot Jupiters around M dwarfs by roughly a factor of 6 compared to isolated single M stars. Relevant works from \cite{Ziegler2020}, \cite{Clark2024} and \cite{Matson2025} all pointed out that the projected separation distribution for M dwarf multiples that are known to host planets peaks beyond 100 AU, indicating that close-in ($<100$~AU) stellar companions probably suppress the formation of planets in M dwarf binaries while wide-orbit stellar companions may not inhibit the process. 

We caution the readers, however, that our field M dwarf sample does not account for metallicity, as the stellar metallicity is not available in the TIC and it necessitates invoking spectroscopic information. Nonetheless, we emphasize that this effect has a minor impact on our finding, and even probably strengthens our conclusions. \cite{Gan2025a} found that giant planets, regardless of their orbital periods, favor metal-rich M dwarfs with a median metallicity [Fe/H] of about 0.3\,dex \citep[see also][]{Dransfield2026} whereas local field M dwarfs are mostly of solar metallicity. With lower metallicities than those M stars hosting hot Jupiters, field M stars are, in principle, more likely to reside in a binary configuration (i.e., higher multiplicity rate) according to observations \citep[e.g.,][]{Moe2019}. This expectation conflicts with the feature we spotted here: a lower multiplicity rate among field M dwarfs compared to those hosting hot Jupiters. However, a few works also argued that stellar multiplicity rate has invariant or even non-monotonic dependence on metallicity when moving beyond 250\,AU \citep{El-Badry2019,Hwang2021,Offner2023} so we are not able to draw a firm conclusion regarding whether metallicity significantly affects our results. A detailed investigation is required but it is beyond the scope of this study. 

The positive influence of wide-orbit stellar companions on the appearance of hot Jupiters around M dwarfs is similar to that of Sun-like star counterparts \citep{Ngo2015,Fontanive2019}. While the exact reason is still under debate, we discuss two possible pathways to produce such a beneficial effect (see also \citealt{Ngo2016}): promoting the planet formation or accelerating the subsequent orbital migration. A higher multiplicity rate might be conducive to hot Jupiter formation by enhancing the formation efficiency via high-density spiral arms induced by stellar companions \citep{Wagner2015,Reynolds2021}. With more solid materials and high gas density within these spiral arms, massive cores or pebbles are more likely to grow there, triggering the birth of large planets. On the other hand, wide-orbit companions could excite the orbits of ``cold'' planets and drive inward migration through the Kozai-Lidov effect \citep{Fabrycky2007,Naoz2011}. The tidal force would then circularize the orbits, leaving behind ``hot'' planets with short orbital periods. Using Equation 42 in \cite{Antognini2015} and assuming that Jupiter-like planets formed at the snow line $r_{\rm snow}=2.7(M_\ast/M_\odot)^{1.75}\ {\rm AU}$ \citep{Ida2005}, and that co-moving stellar companions have circular orbits, we perform a rough estimation of the Kozai-Lidov effect timescales $\tau_{\rm KL}$ of 13 systems that have wide-orbit stellar companions. The masses of stellar companions are estimated through the Gaia $G$ band magnitude difference with respect to the host star \citep{Mamajek2008}, except for \tar B that we have the mass determined. We find that $\tau_{\rm KL}$ ranges from about $10^{-3}$ Gyr to 100 Gyr, with 9 systems (69\%) having $\tau_{\rm KL}$ smaller than 10 Gyr, indicating that the Kozai–Lidov mechanism is dynamically efficient during the orbital evolution in the majority of these systems. Nevertheless, we note that the assumptions we used above are simplified and might not be realistic, which require further investigations. In either scenario, our findings, together with results on Sun-like stars, provide a unique constraint that should be reproduced by planet formation simulations. Stellar obliquity studies on both hot and warm Jupiters around M dwarfs \citep[e.g.,][]{Gan2024} will offer a valuable test to these competing theories. 

\section{Conclusions}\label{sec:conclusion}

In this work, we report the discovery and characterization of \tar Ab, a giant planet transiting a mid-M dwarf ($M_\ast=0.36\pm0.02\ M_\odot$) every 4.34 days, accompanied by a wide-orbit white dwarf with a mass of $0.59\pm0.16\ M_\odot$ at a projected separation about 2,500 AU. The planet \tar Ab has a radius of $0.74\pm0.04\ R_J$ and a mass of $0.09\pm0.04\ M_J$ with a $3\sigma$ upper limit of $0.22\ M_J$. Building on this system, we conduct a systematic search for co-moving stellar companions at projected distances between 100 and 10,000\,AU around all M dwarfs hosting confirmed giant planets with orbital period shorter than 10 days and radii greater than $0.7\ R_J$. We measure a multiplicity rate of $P(\rm Comp|HJ)=34.2\pm9.5\%$ for the planet sample, which is significantly higher than the value of field M stars with $P(\rm Comp)=5.3\pm3.7\%$ by approximately a factor of 6. Our results indicate that a wide-orbit stellar companion between 100 and 10,000\,AU can facilitate the growth of short-period giant planets around M dwarfs with masses $0.21\leq M_\ast\leq0.64\ M_\odot$, and high-eccentricity migration may be an important route to produce such systems. 

\section{Acknowledgments}

T.G. and S.M. acknowledge support by the National Natural Science Foundation of China (No. 12133005). T.G. and E.P. acknowledge financial support from the Agencia Estatal de Investigaci\'on of the Ministerio de Ciencia e Innovaci\'on MCIN/AEI/10.13039/501100011033 and the ERDF “A way of making Europe” through projects PID2021-125627OB-C32 and PID2024-158486OB-C32, and the funding from the European Union (ERC AdvG SPEAR, GA 101200674). Funding for K.B. was provided by the European Union (ERC AdG SUBSTELLAR, GA 101054354). G.D. acknowledges funding from Magdalen College, Oxford. This work is partly supported by JSPS KAKENHI Grant Numbers  JP24H00017, JP25K24620, JP26H01402, JP26K00755, and JP24K00689. A.L., É.A., C.C., R.D. and N.J.C. acknowledge the financial support of the Fonds de recherche du Québec - Secteur Nature et technologies (FRQ-NT) through the Centre de recherche en astrophysique du Québec as well as the support from the Trottier Family Foundation and the Trottier Institute for Research on Exoplanets. A.L. acknowledges support from the FRQ-NT under file \#349961. É.A. and R.D. acknowledge support from Canada Foundation for Innovation (CFI) program, the Université de Montréal and Université Laval, the Canada Economic Development (CED) program and the Ministere of Economy, Innovation and Energy (MEIE). S.M.G. acknowledges the support of the Natural Sciences and Engineering Research Council of Canada
(NSERC) and is partially funded through an NSERC Postdoctoral Fellowship (PDF). KAC acknowledges support from the TESS mission via subaward s3449 from MIT and NASA grants 80NSSC24K1889 and 80NSSC26K0081.

This research uses data obtained through the Telescope Access Program (TAP), which has been funded by the TAP association, including Center for Astronomical Mega-Science CAS(CAMS), PKU, THU, USTC, and WLU.

Based on observations obtained at the Canada-France-Hawaii Telescope (CFHT) which is operated from the summit of Maunakea by the National Research Council of Canada, the Institut National des Sciences de l'Univers of the Centre National de la Recherche Scientifique of France, and the University of Hawaii. The observations at the Canada-France-Hawaii Telescope were performed with care and respect from the summit of Maunakea which is a significant cultural and historic site. Based on observations obtained with SPIRou, an international project led by Institut de Recherche en Astrophysique et Planétologie, Toulouse, France.

Funding for the TESS mission is provided by NASA's Science Mission Directorate. We acknowledge the use of public TESS data from pipelines at the TESS Science Office and at the TESS Science Processing Operations Center. Resources supporting this work were provided by the NASA High-End Computing (HEC) Program through the NASA Advanced Supercomputing (NAS) Division at Ames Research Center for the production of the SPOC data products. This research has made use of the Exoplanet Follow-up Observation Program website, which is operated by the California Institute of Technology, under contract with the National Aeronautics and Space Administration under the Exoplanet Exploration Program. This paper includes data collected by the \tess\ mission, which are publicly available from the Mikulski Archive for Space Telescopes\ (MAST). All 2-min TESS data and Target Pixel Files used in this work can be accessed via MAST: \dataset[10.17909/t9-nmc8-f686]{https://doi.org/10.17909/t9-nmc8-f686} and \dataset[10.17909/t9-yk4w-zc73]{https://doi.org/10.17909/t9-yk4w-zc73}.

This work has made use of data from the European Space Agency (ESA) mission
{\it Gaia} (\url{https://www.cosmos.esa.int/gaia}), processed by the {\it Gaia} Data Processing and Analysis Consortium (DPAC,
\url{https://www.cosmos.esa.int/web/gaia/dpac/consortium}). Funding for the DPAC has been provided by national institutions, in particular the institutions participating in the {\it Gaia} Multilateral Agreement.

This article is based on observations made with the MuSCAT2 instrument, developed by ABC, at Telescopio Carlos Sánchez operated on the island of Tenerife by the IAC in the Spanish Observatorio del Teide. This work is partly supported by JSPS KAKENHI Grant Numbers JP18H01265 and JP18H05439, and JST PRESTO Grant Number JPMJPR1775.


This work makes use of observations from the LCOGT network. Part of the LCOGT telescope time was granted by NOIRLab through the Mid-Scale Innovations Program (MSIP), which is funded by NSF.

TRAPPIST is funded by the Belgian Fund for Scientific Research (Fond National de la Recherche Scientifique, FNRS) under the grant FRFC 2.5.594.09.F, with the participation of the Swiss National Science Fundation (SNF). M.G. and E.J. are F.R.S.-FNRS Research Directors.

The ULiege's contribution to SPECULOOS has received funding from the European Research Council under the European Union's Seventh Framework Programme (FP/2007-2013) (grant Agreement n$^\circ$ 336480/SPECULOOS), from the Balzan Prize and Francqui Foundations, from the Belgian Scientific Research Foundation (F.R.S.-FNRS; grant n$^\circ$ T.0109.20), from the University of Liege, and from the ARC grant for Concerted Research Actions financed by the Wallonia-Brussels Federation. The Cambridge contribution is supported by a grant from the Simons Foundation (PI Queloz, grant number 327127). J.d.W. and MIT gratefully acknowledge financial support from the Heising-Simons Foundation, Dr. and Mrs. Colin Masson and Dr. Peter A. Gilman for Artemis, the first telescope of the SPECULOOS North network situated in Tenerife, Spain. The Birmingham contribution to SPECULOOS has received fund from the European Research Council (ERC) under the European Union's Horizon 2020 research and innovation programme (grant agreement n$^\circ$ 803193/BEBOP), from the MERAC foundation, and from the Science and Technology Facilities Council (STFC; grant n$^\circ$ ST/S00193X/1, ST/W002582/1, and ST/Y001710/1) and from the ERC/UKRI Frontier Research Guarantee programme (EP/Z000327/1/CandY).

This material is based upon work supported by the National Aeronautics and Space Administration under Agreement No.\ 80NSSC21K0593 for the program ``Alien Earths.''
The results reported herein benefited from collaborations and/or information exchange within NASA's Nexus for Exoplanet System Science (NExSS) research coordination network sponsored by NASA's Science Mission Directorate. Visiting Astronomer at the Infrared Telescope Facility, which is operated by the University of Hawaii under contract 80HQTR24DA010 with the National Aeronautics and Space Administration.

%

\vspace{5mm}
\facilities{TESS, Gaia, CFHT/SPIRou, IRTF/SpeX, Shane/Kast, Palomar/PHARO, Gemini-N/`Alopeke, LCOGT, SPECULOOS, TRAPPIST-North, MuSCAT2, Palomar/WIRC}

\software{AstroImageJ \citep{Collins2017}, astroquery \citep{Ginsburg2019}, astropy \citep{Astropy2013,Astropy2018,Astropy2022}, juliet \citep{juliet}, batman \citep{Kreidberg2015}, radvel \citep{Fulton2018}}





\appendix

\counterwithin{table}{section}
\counterwithin{figure}{section}

\section{Ground-based Observations}\label{ground_based_observations}

\subsection{Photometric Observations}

\subsubsection{LCOGT}

We collected three transits of \tar Ab using 1-meter telescopes from the Las Cumbres Observatory Global Telescope network \citep[LCOGT;][]{Brown2013} on UT 2022 June 9, UT 2023 January 25 and UT 2023 March 19. We carried out observations in $z'$, $r'$ and $g'$ band from Teide Observatory, South African Astronomical Observatory (SAAO) and Cerro Tololo Interamerican Observatory (CTIO) with exposure times of 500s, 500s and 600s, respectively. All three observations were conducted with the Sinistro cameras, with a field of view (FOV) of $26'\times26'$ and a plate scale of $0.389\arcsec$/pixel. Except for the second observation that was stopped before the egress, the other two runs covered the full transit event. The raw data were calibrated with the \code{BANZAI} pipeline \citep{McCully2018}, and we then employed the \code{AstroImageJ} software \citep{Collins2017} to perform the photometric analysis on the reduced images. We confirmed the achomatic transit signal on the target.

\subsubsection{SPECULOOS}


We used the SPECULOOS \citep[Search for habitable Planets EClipsing ULtra-cOOl Stars,][]{ Delrez2018,Sebastian_2021AA,Burdanov2022} 1.0m network to observe four full transits and one partial transit of TOI-5628A\,b. Each telescope equipped with a 2K$\times$2K Andor iKon-L camera with a pixel scale of $0.35\arcsec$ and a FOV of $12\arcmin\times12\arcmin$.
Four transits were observed with SPECULOOS-South located in Paranal, Chile and one transit was observed with SPECULOOS-North at Teide Observatory.
First transit was carried out in the Sloan-$g'$ filter on UT 2023 January 25. Second and third transits were carried out on UT 2023 March 31 in the Sloan-$z'$ and Sloan-$g'$ filters.
Fourth and fifth transits were observed on UT 2023 May 31 in the Sloan-$r'$ and Sloan-$i'$ filters. The data reduction and photometric extraction were performed using the {\tt PROSE} pipeline \citep{prose_2022}. The analysis of the resulting light curves did not reveal any transit depth chromaticity.

\subsubsection{TRAPPIST-North}


TRAPPIST-North \citep{Barkaoui2019_TN} is a 60-cm Ritchey-Chr\'etien telescope located at Oukaimeden Observatory, and it is equipped with a thermoelectrically cooled 2K$\times$2K Andor iKon-L BEX2-DD CCD camera with a pixel scale of 0.6\arcsec, and a FOV of $20\arcmin\times20\arcmin$. The TRAPPIST-North telescope observed a full transit of \tar Ab in the $I+z$ filter on UT 2023 March 18 with an exposure time of 100s. Data processing and photometric extraction were done using the {\tt PROSE} pipeline \citep{prose_2022}. The photometric measurements were extracted using an uncontaminated target aperture of 4.8\arcsec.

\subsubsection{MuSCAT2}


We acquired a full transit of \tar Ab on the night of UT 2024 March 30 using the multicolor imager MuSCAT2 \citep{2019JATIS...5a5001N} mounted on the 1.52 m Telescopio Carlos S\'{a}nchez at Teide Observatory in Spain. MuSCAT2 has a field of view of $7.4' \times 7.4'$ with a pixel scale of $0.44''$ pixel$^{-1}$, and it allows to conduct observations simultaneously in four bands ($g$, $r$, $i$, and $z_s$). After standard dark and flat calibration, the photometric analysis was performed using MuSCAT2 pipeline (\citealp{Parviainen2019}), which optimized the aperture to minimize the photometric dispersion and modeled the light curve with instrumental systematic effects included. The exposure times were set to 60s in $g$ and $r$, 15s in $i$, and 30s in $z_s$. We excluded the $g$ band data due to its low SNR and large scatter.

\subsubsection{WIRC}

We observed one transit of \tar\ in the $J$ band on UT 2024 April 17 using the Wide-field InfraRed Camera (WIRC; \citealt{Wilson2003, 2019PASP..131b5001T}) at the prime focus of the 200-inch Hale Telescope at Palomar Observatory with an exposure time of 20s. The airmass varied from 1.03 to 1.41 across the observation. The calibration and photometry analysis were perform using \code{exowirc} \citep{2020AJ....159..108V} and \code{photutils} \citep{2016ascl.soft09011B}, with an optimized aperture radius that minimizing the scatter of the light curve. We cleaned the light curve through a pre-fit using a median-filtered light curve, followed by a sigma-clipping, resulting in 517 points spanning the transit.

\begin{figure*}
\centering
\includegraphics[width=0.99\textwidth]{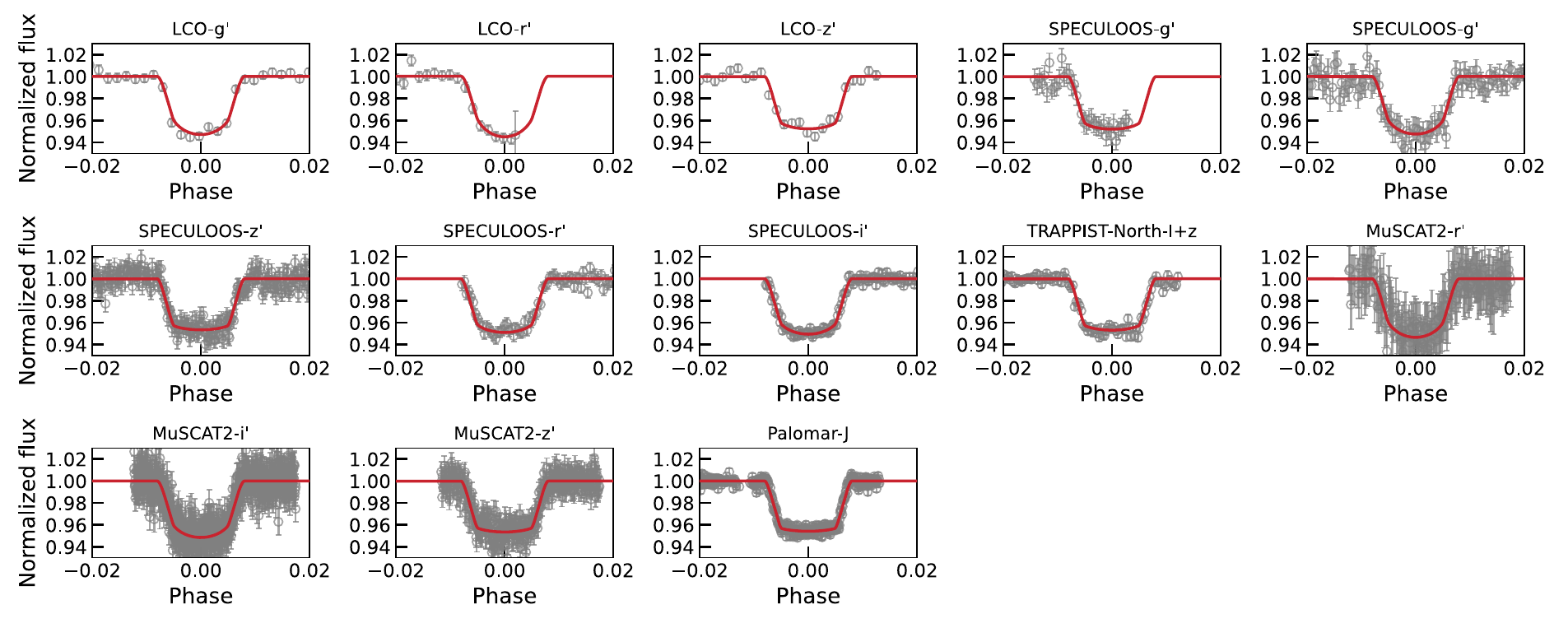}
\caption{Phase-folded ground-based photometric observations of \tar Ab along with the best-fit transit models shown in red from the joint analysis. }
\label{groundfit}
\end{figure*}

\subsection{Spectroscopic Observations}

\subsubsection{IRTF/SpeX observation on \tar A}

\begin{figure}
\centering
\includegraphics[width=0.49\textwidth]{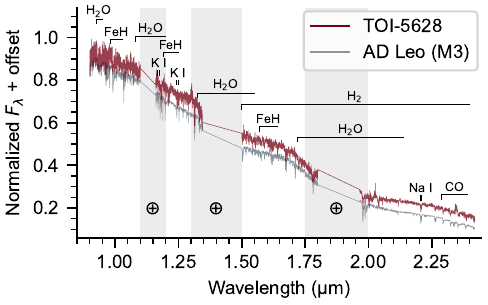}
\includegraphics[width=0.49\textwidth]{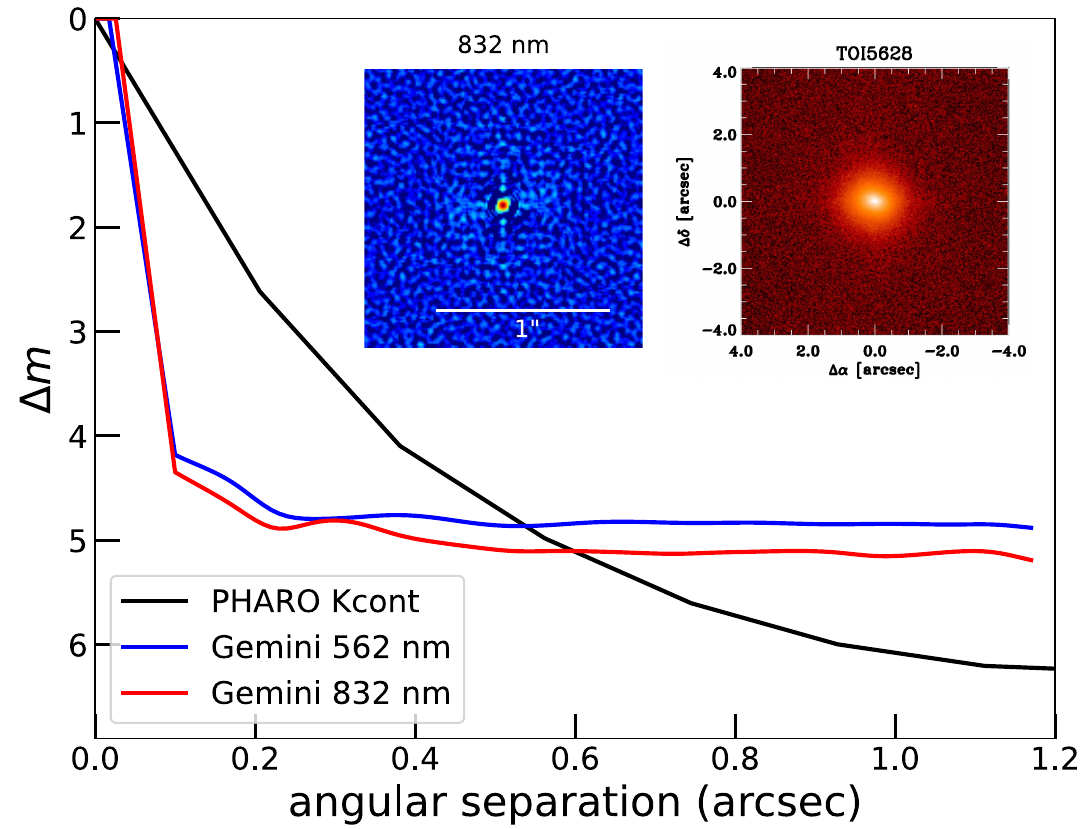}
\caption{Left panel: SpeX SXD spectrum of \tar A (red) and the comparison spectrum of an M3V dwarf AD Leo (gray) from the IRTF Spectral Library \citep{Rayner2009}. Strong atomic and molecular features are marked. Right panel: $5\sigma$ contrast curves for \tar A from Palomar/PHARO and Gemini-N/`Alopeke with images shown as inset figures. The co-moving white dwarf companion \tar B at about $23''$ is outside the FOV of both instruments, and there is no additional close stellar companions within the detection limits.}
\label{SpeX_imaging}
\end{figure}

We obtained a near-infrared spectrum of TOI-5628A with the SpeX spectrograph \citep{Rayner2003} on the 3.2-m NASA Infrared Telescope Facility (IRTF) on UT 2022 July 21.
We used the short-wavelength cross-dispersed (SXD) mode with the $0\farcs3 \times 15''$ slit aligned to the parallactic angle, yielding spectra spanning 0.82--2.42\,$\micron$ at a resolving power of $R{\sim}2000$.
Conditions were clear with seeing of $0\farcs9$.
We collected six 300-s exposures at an average airmass of 1.7, nodding along the slit in an ABBA pattern, followed by the standard set of SXD flat-field and arc-lamps calibrations and six 10-s exposures of the A0\,V telluric standard at a similar airmass.
The data were reduced with Spextool v4.1 \citep{Cushing2004}, following a standard approach detailed previously \citep{Barkaoui2025, Ghachoui2024}.
The final spectrum, shown in the left panel of Fig.\,\ref{SpeX_imaging}, has a median SNR of 79 per resolution element.

\subsubsection{Shane/Kast observation on \tar B} \label{sec:shane}

We observed the spectrum of \tar B with the Kast Double Spectrograph on the Shane 3m telescope at Lick Observatory on the night of UT 2022 June 27. Given the faintness of the star, we adopted exposure times of 3000s and 1500s for the blue (3700-5400~$\AA$) and red (5900-9000~$\AA$) bands, respectively. Data were reduced using the \code{kastredux} package \citep{adam_burgasser_2026_18333308}. The reduced spectrum is shown in the bottom panel of Figure~\ref{fig:sed}.


\subsection{High Angular Resolution Imaging Observations}

\subsubsection{Palomar/PHARO}

We observed \tar A on UT 2024 February 15 with the Palomar High Angular Resolution Observer \citep[PHARO;][]{Hayward2001}, which is an adaptive optics imaging instrument having a plate scale of 0.025 arcsec pixel$^{-1}$ mounted on the 5-m Hale Telescope. The observation was performed in $K$cont filter with an estimated point spread function (PSF) of 0.18 arcsec, reaching a contrast of about 4.981 mag at 0.5 arcsec. The data were reduced with the pipeline introduced in \citep{Furlan2017}. No nearby sources were identified within the detection limits achieved (see Figure~\ref{SpeX_imaging}). 

\subsubsection{Gemini-N/`Alopeke}
\tar A was also observed on UT 2024 May 24 using the `Alopeke speckle instrument on the Gemini North 8-m telescope \citep{Scott2021,Howell2022}.  `Alopeke provides simultaneous speckle imaging in two bands (562 nm and 832 nm) with output data products including a reconstructed image with robust contrast limits on companion detections. Twelve sets of $1000\times0.06$ second images were obtained and processed in our standard reduction pipeline \citep{Howell2011}. Figure~\ref{SpeX_imaging} shows our final contrast curves and the 832 nm reconstructed speckle image. We find that TOI-5628A does not have any companion brighter than 5 magnitudes below that of the target star from the 8-m telescope diffraction limit (20 mas) out to $1.2''$. At the distance of \tar A (d=110 pc), these angular limits correspond to spatial limits of 2.2 to 132 AU.

\section{The SEDs of \tar A and \tar B, and the Shane/Kast spectrum of \tar B}

\begin{figure*}
    \centering
    \includegraphics[width=0.95\textwidth]{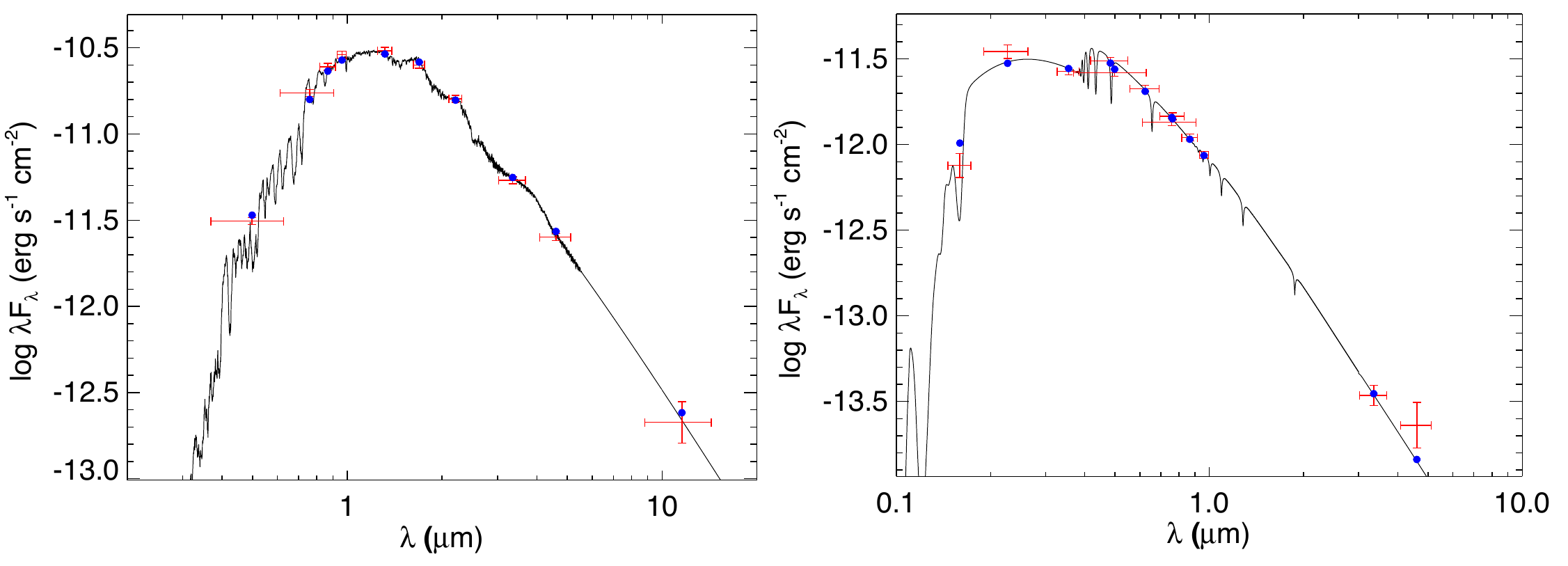}
    \includegraphics[width=0.93\textwidth]{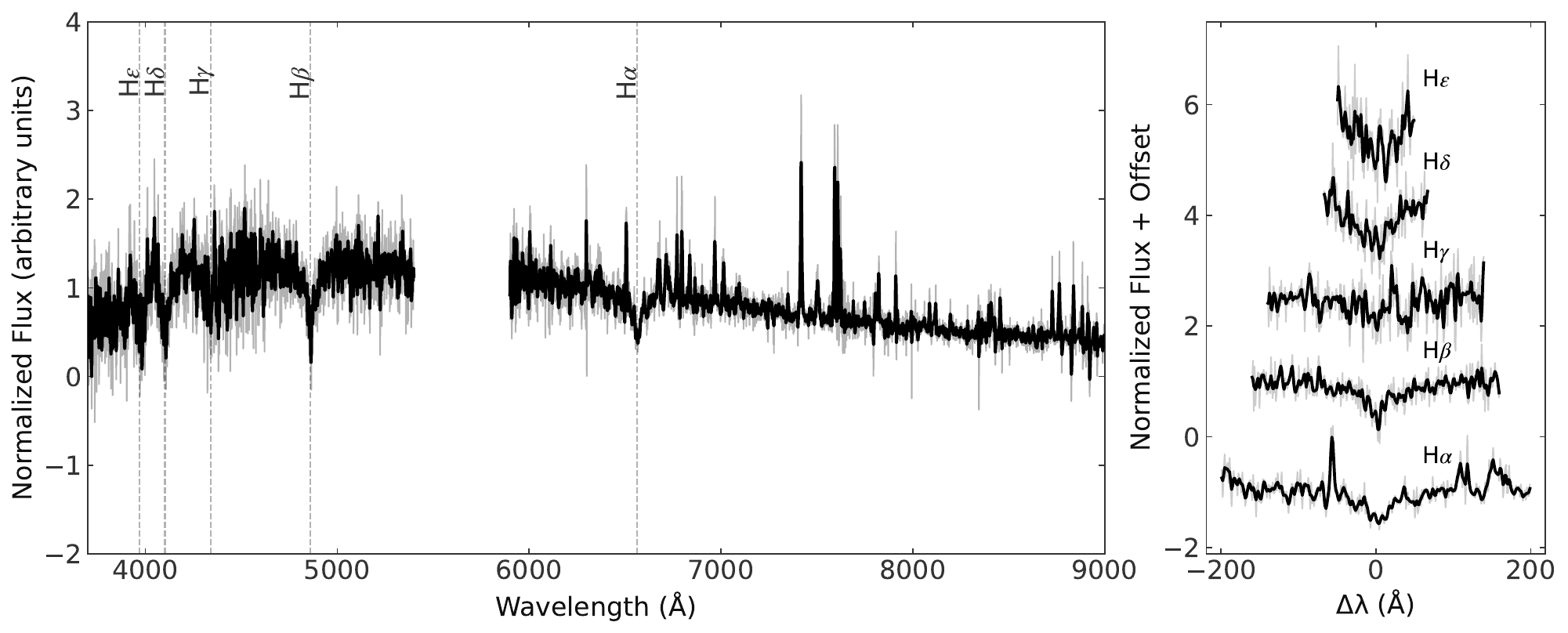}
\caption{Spectral energy distribution of TOI-5628A (Top left) and the associated white dwarf TOI-5628B (Top right). Red symbols represent the observed photometric measurements, where the horizontal bars represent the effective width of the passband. Blue dots are the model fluxes from the best-fit PHOENIX (left) and \citealt{Koester2010} (right) atmosphere models (black). Bottom left: The stitched blue ($3700-5400 \AA$) and red ($5900-9000 \AA$) Shane/Kast spectrum of \tar B, normalized to an arbitrary continuum level. The raw spectrum is shown in gray, whereas a Gaussian-smoothed spectrum ($\sigma = 1\AA$) is plotted in black. Clear Balmer absorption features are marked with dashed vertical lines. Bottom right: Normalized Balmer line profiles plotted as a function of $\Delta \lambda$ from line center. Each line is vertically offset for clarity. These spectral features confirm the DA WD nature of \tar B.}
\label{fig:sed}
\end{figure*}

The top panels of Figure~\ref{fig:sed} present the broadband photometry of \tar A and \tar B along with the best-fit SED models. The bottom panel of Figure~\ref{fig:sed} shows the Shane/Kast spectrum of \tar B.

\section{The ZTF light curve of \tar A}

\begin{figure*}
\centering
\includegraphics[width=0.99\textwidth]{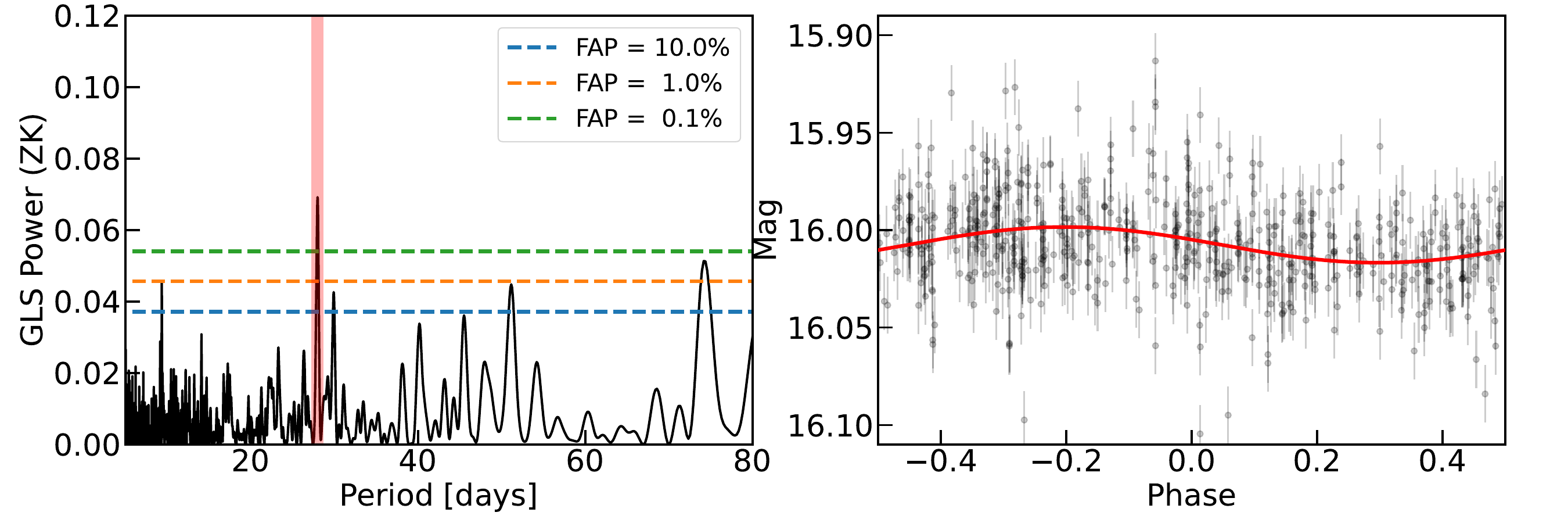}
\caption{Left panel: The GLS periodogram of the ZTF light curve. The vertical red line represents the rotational period inferred for \tar A. The false alarm probability levels of 10\%, 1\%, and 0.1\% are shown as horizontal dashed lines with different colors. Right panel: Phase-folded ZTF light curve at 27.9 days with the best-fit sinusoidal model, shown as a red solid curve.}
\label{ZTF}
\end{figure*}

Figure~\ref{ZTF} shows the long-term light curve of \tar A from ZTF, where a $27.9$-day photometric modulation can be seen in the data.

\bibliography{planet}{}
\bibliographystyle{aasjournal}



\end{document}